\documentclass[lettersize,journal]{IEEEtran}
\usepackage{amsmath,amsfonts}
\usepackage{algorithmic}
\usepackage{algorithm}
\usepackage{array}
\usepackage[caption=true,font=small]{subfig}
\usepackage{balance}
\usepackage{textcomp}
\usepackage{stfloats}
\usepackage{url}
\usepackage{verbatim}
\usepackage{graphicx}
\graphicspath{{./eps}{./png}}
\usepackage{cite}
\usepackage{tikz}
\usetikzlibrary{arrows.meta}
\usepackage{comment}
\usepackage{balance}
\usepackage{hyperref}
\usepackage{algorithmic}
\usepackage{algorithm}

\usepackage{epsfig}
\usepackage{mathrsfs}
\usepackage{mdframed}
\usepackage[acronym]{glossaries}
\usepackage{textcomp}
\usepackage{amssymb}% http://ctan.org/pkg/amssymb
\usepackage{pifont}% http://ctan.org/pkg/pifont
\newcommand{\cmark}{\ding{51}}%
\newcommand{\xmark}{\ding{55}}%

\newtheorem{remark}{Remark}

\newtheorem{problem}{Problem}

\newtheorem{definition}{Definition}

\usepackage{epstopdf}

\newcommand{\sseq}{\mathbf{s}}
\newcommand{\wseq}{\mathbf{w}}
\newcommand{\useq}{\mathbf{u}}
\newcommand{\yseq}{\mathbf{y}}
\newcommand{\xseq}{\mathbf{x}}

\newcommand{\E}{\mathsf{E}}
\newcommand{\qed}{$\hfill\square$}

\newacronym{ev}{EV}{ego vehicle}
\newacronym{ov}{OV}{opponent vehicle}
\newacronym{mpc}{MPC}{Model Predictive Control}
\newacronym{mip}{MIP}{Mixed-Integer Planning}
\newacronym{ltl}{LTL}{Linear-time Temporal Logic}
\newacronym{stl}{STL}{Signal Temporal Logic}
\newacronym{dtl}{DTL}{Distribution Temporal Logic}
\newacronym{dstl}{DSTL}{Distribution Signal Temporal Logic}
\newacronym{mdp}{MDP}{Markov decision process}
\newacronym{momdp}{MOMDP}{mixed observable Markov decision process}
\newacronym{pomdp}{POMDP}{partially observable Markov decision process}
\newacronym{poctl}{POCTL}{partially observable computation tree logic}
\newacronym{pce}{PCE}{Polynomial Chaos Expansion}
\newacronym{gmm}{GMM}{Gaussian mixture model}
\newacronym{hmm}{HMM}{hidden Markov model}
\newacronym{pdf}{PDF}{probabilistic density function}

\begin{document}

\title{\LARGE \bf Intention-Aware Control Based on Belief-Space Specifications and
Stochastic Expansion}

\author{Zengjie Zhang,~\IEEEmembership{Member,~IEEE,}
        Zhiyong Sun,~\IEEEmembership{Member,~IEEE,}
        and Sofie Haesaert,~\IEEEmembership{Member,~IEEE,}
        % <-this % stops a space
\thanks{This work was supported by the European project SymAware under grant No. 101070802.}% <-this % stops a space
\thanks{$^{1}$Zengjie Zhang, Zhiyong Sun, and Sofie Haesaert are with the Department of Electrical Engineering, Eindhoven University of Technology,
        PO Box 513, 5600 MB Eindhoven, Netherlands.
        {\tt\small \{z.zhang3, z.sun, s.haesaert\}@tue.nl}} \\
        \thanks{$^{2}$Zhiyong Sun is also with the College of Engineering, Peking University, Beijing, China. {\tt\small \{zhiyong.sun@pku.edu.cn\}}}}

% The paper headers
\markboth{Journal of \LaTeX\ Class Files,~Vol.~14, No.~8, August~2021}%
{Shell \MakeLowercase{\textit{et al.}}: A Sample Article Using IEEEtran.cls for IEEE Journals}

% \IEEEpubid{0000--0000/00\$00.00~\copyright~2021 IEEE}
% Remember, if you use this you must call \IEEEpubidadjcol in the second
% column for its text to clear the IEEEpubid mark.

\maketitle

\begin{abstract}
This paper develops a correct-by-design controller for an autonomous vehicle interacting with opponent vehicles with unknown intentions. We define an intention-aware control problem incorporating epistemic uncertainties of the opponent vehicles and model their intentions as discrete-valued random variables. Then, we focus on a control objective specified as belief-space temporal logic specifications. From this stochastic control problem, we derive a sound deterministic control problem using stochastic expansion and solve it using shrinking-horizon model predictive control. The solved intention-aware controller allows a vehicle to adjust its behaviors according to its opponents' intentions. It ensures provable safety by restricting the probabilistic risk under a desired level. We show with experimental studies that the proposed method ensures strict limitation of risk probabilities, validating its efficacy in autonomous driving cases. This work provides a novel solution for the risk-aware control of interactive vehicles with formal safety guarantees.
\end{abstract}


\section{Introduction}\label{sec:intro}
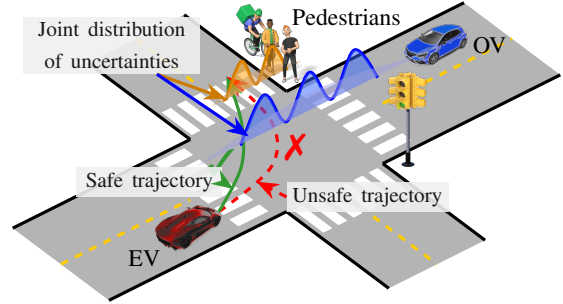
\begin{figure}[htbp]
\noindent
\hspace*{\fill} 
\begin{tikzpicture}[scale=1]

\definecolor{vvgray}{RGB}{240, 240, 240}
\definecolor{vsgray}{RGB}{180, 180, 180}
\definecolor{vsyellow}{RGB}{255, 204, 0}
\definecolor{sblue}{RGB}{51, 100, 255}
\definecolor{ssblue}{RGB}{159, 159, 223}
\definecolor{syellow}{RGB}{255, 153, 0}
\definecolor{dblue}{RGB}{33, 33, 95}
\definecolor{sorg}{RGB}{204, 154, 102}
\definecolor{dorg}{RGB}{102, 51, 0}
\definecolor{dyellow}{RGB}{204, 122, 0}
\definecolor{dgreen}{RGB}{51, 153, 51}

\def\x{0.32}

\path[fill=vsgray] (-2-2*\x, 1.5-\x) -- (-2*\x-2*\x, -\x+1.5*\x) --
(-3-2*\x, -1.5+1.5*\x) -- (-3+2*\x, -1.5-1.5*\x) -- 
(-2*\x+2*\x, -\x-1.5*\x) -- (2-2*\x, -1.5-\x) --
(2+2*\x, -1.5+\x) -- (2*\x+2*\x, \x-1.5*\x) --
(3+2*\x, 1.5-1.5*\x) -- (3-2*\x, 1.5+1.5*\x) --
(2*\x-2*\x, \x+1.5*\x) -- (-2+2*\x, 1.5+\x) -- cycle;

\draw [vsyellow, very thick, dashed] (-2, 1.5) -- (-1.8*2*\x, 1.8*1.5*\x);
\draw [vsyellow, very thick, dashed] (1.8*2*\x, -1.8*1.5*\x) -- (2, -1.5);

\draw [black, very thick] (-2-2*\x, 1.5-\x) -- (-2*\x-2*\x, -\x+1.5*\x);
\draw [black, very thick] (-2*\x+2*\x, -\x-1.5*\x) -- (2-2*\x, -1.5-\x);

\draw [black, very thick] (-2+2*\x, 1.5+\x) -- (2*\x-2*\x, \x+1.5*\x);
\draw [black, very thick] (2*\x+2*\x, \x-1.5*\x) -- (2+2*\x, -1.5+\x);

\draw [vsyellow, very thick, dashed] (-3, -1.5) -- (-2*2*\x, -2*\x);
\draw [vsyellow, very thick, dashed] (2*2*\x, 2*\x) -- (3, 1.5);

\draw [black, very thick] (-3-2*\x, -1.5+1.5*\x) -- (-2*\x-2*\x, -\x+1.5*\x);
\draw [black, very thick] (2*\x-2*\x, \x+1.5*\x) -- (3-2*\x, 1.5+1.5*\x);

\draw [black, very thick] (-3+2*\x, -1.5-1.5*\x) -- (-2*\x+2*\x, -\x-1.5*\x);
\draw [black, very thick] (2*\x+2*\x, \x-1.5*\x) -- (3+2*\x, 1.5-1.5*\x);

\def\l{0.1}
\def\ll{0.03}

\foreach \i in {-0.1, -0.2, -0.3, -0.4, -0.5, -0.6}
        \path[fill=white] (2*\i-2*\l-2*\ll-0.2, \i+1.5*\l-\ll+1) -- (2*\i-2*\l+2*\ll-0.2, \i+1.5*\l+\ll+1) -- (2*\i+2*\l+2*\ll-0.2, \i-1.5*\l+\ll+1)-- (2*\i+2*\l-2*\ll-0.2, \i-1.5*\l-\ll+1) --cycle;
        
\foreach \i in {-0.1, -0.2, -0.3, -0.4, -0.5, -0.6}
        \path[fill=white] (2*\i-2*\l-2*\ll+1.55, \i+1.5*\l-\ll-0.3) -- (2*\i-2*\l+2*\ll+1.55, \i+1.5*\l+\ll-0.3) -- (2*\i+2*\l+2*\ll+1.55, \i-1.5*\l+\ll-0.3)-- (2*\i+2*\l-2*\ll+1.55, \i-1.5*\l-\ll-0.3) --cycle;
        
\foreach \i in {-0.1, -0.2, -0.3, -0.4, -0.5, -0.6}
        \path[fill=white] (2*\i-2*\l+2*\ll+1.6, -1.5*\i-1*\l-1.5*\ll-0.05) -- (2*\i-2*\l-2*\ll+1.6, -1.5*\i-1*\l+1.5*\ll-0.05) -- (2*\i+2*\l-2*\ll+1.6, -1.5*\i+1*\l+1.5*\ll-0.05)-- (2*\i+2*\l+2*\ll+1.6, -1.5*\i+1*\l-1.5*\ll-0.05) --cycle;

\foreach \i in {-0.1, -0.2, -0.3, -0.4, -0.5, -0.6}
        \path[fill=white] (2*\i-2*\l+2*\ll-0.3, -1.5*\i-1*\l-1.5*\ll-1) -- (2*\i-2*\l-2*\ll-0.3, -1.5*\i-1*\l+1.5*\ll-1) -- (2*\i+2*\l-2*\ll-0.3, -1.5*\i+1*\l+1.5*\ll-1)-- (2*\i+2*\l+2*\ll-0.3, -1.5*\i+1*\l-1.5*\ll-1) --cycle;   

\node[inner sep=2pt, anchor=center] (ped3) at (-0.3, 1.6) {\includegraphics[width=1cm]{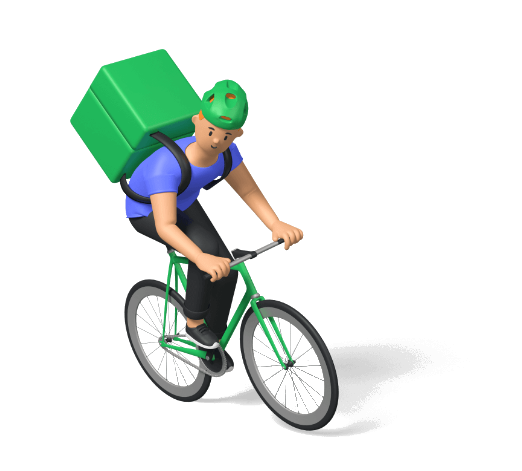}};
\node[inner sep=2pt, anchor=center] (ped2) at (-0.1, 1.4) {\includegraphics[width=0.8cm]{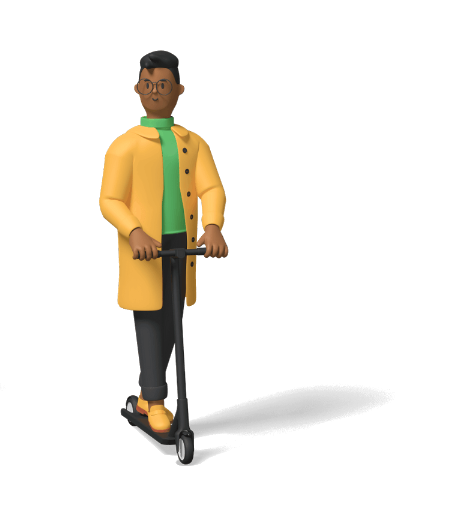}};
\node[inner sep=2pt, anchor=south west] () at (0, 1.6) {\small Pedestrians};

\node[inner sep=2pt, anchor=center] (ego) at (-1.3, -1) {\includegraphics[width=1cm]{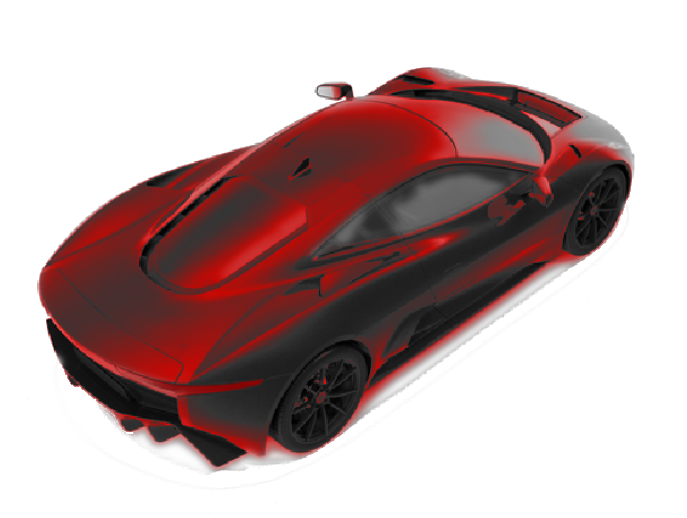}};
\node[inner sep=2pt, anchor=south east] () at (-1.6, -1.6) {\small EV};

\node[inner sep=2pt, anchor=center] (oppo) at (2, 1.4){\includegraphics[width=1cm]{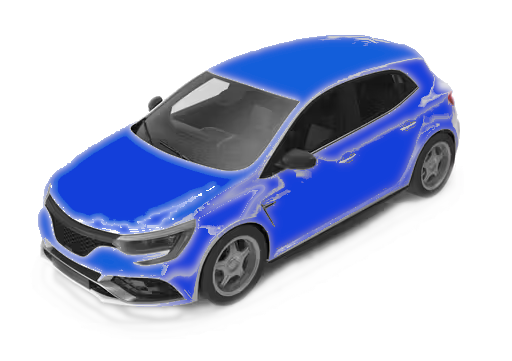}};
\node[inner sep=2pt, anchor=north east] () at (3.0, 1.6) {\small OV};

\node[inner sep=2pt, anchor=center] (tf) at (1.6, 0.4) {\includegraphics[width=1.5cm]{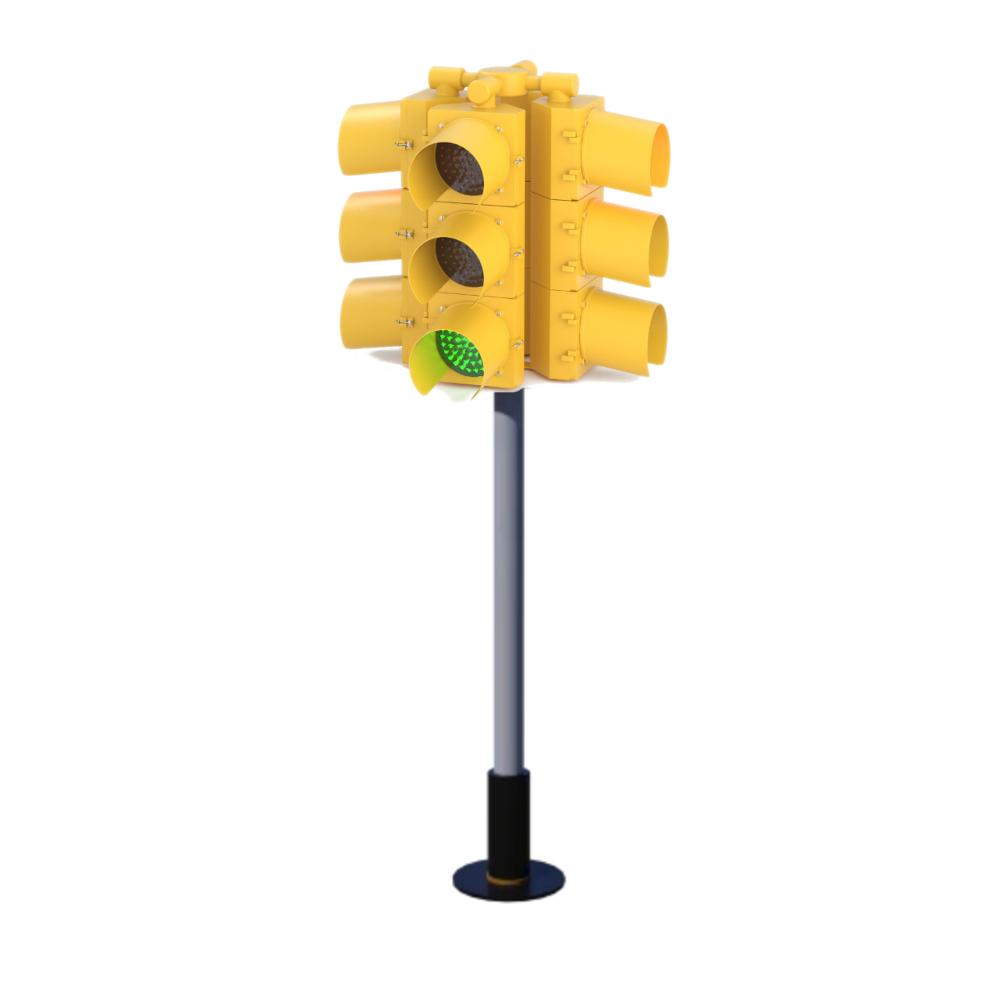}};

\draw [->, >=Stealth, very thick, red, dashed] plot [smooth, tension=0.6] coordinates { (-0.9, -0.8) (-0.2, -0.2) (-0.2, 0.4) (-0.8, 1)};
\node[inner sep=2pt, anchor=south west, red] () at (-0.1, -0.2){\Large \textbf{\xmark}};

\draw [->, >=Stealth, very thick, dgreen] plot [smooth, tension=0.6] coordinates { (-0.9, -0.8) (-0.6, -0.2) (-0.6, 0.4) (-0.8, 1)};
\node[inner sep=2pt, anchor=south east, dgreen] () at (-0.6, -0.4){\Large \textbf{\cmark}};

\def\ofc{1.1}
\def\ofd{1.5}

\draw [very thick, dyellow] (-0.05, -0.05/2+\ofc) -- plot [smooth, tension=0.6] coordinates { (0.0, 0.0/2+\ofc) (-0.2, -0.2/2+\ofd) (-0.4, -0.4/2+\ofc) (-0.6, -0.6/2+\ofd) (-0.8, -0.8/2+\ofc)} -- (-0.85, -0.85/2+\ofc);
\path[fill=syellow,opacity=0.5] (-0.05, -0.05/2+\ofc) -- plot [smooth, tension=0.6] coordinates { (0.0, 0.0/2+\ofc) (-0.2, -0.2/2+\ofd) (-0.4, -0.4/2+\ofc) (-0.6, -0.6/2+\ofd) (-0.8, -0.8/2+\ofc)} -- (-0.85, -0.85/2+\ofc) -- cycle;

%\draw [very thick, dorg, dotted] (0.25, 0.25/2+\ofc) -- plot [smooth, tension=0.6] coordinates { (0.2, 0.2/2+\ofc)  (-0.4, -0.4/2+0.85*\ofd) (-1.0, -1.0/2+\ofc)} -- (-1.05, -1.05/2+\ofc);
\path[fill=sorg,opacity=0.5] (0.25, 0.25/2+\ofc) -- plot [smooth, tension=0.6] coordinates { (0.2, 0.2/2+\ofc)  (-0.4, -0.4/2+0.85*\ofd) (-1.0, -1.0/2+\ofc)} -- (-1.05, -1.05/2+\ofc) -- cycle;

\node[inner sep=2pt, anchor=center] (ped1) at (0.2cm, 1.3cm) {\includegraphics[width=0.7cm]{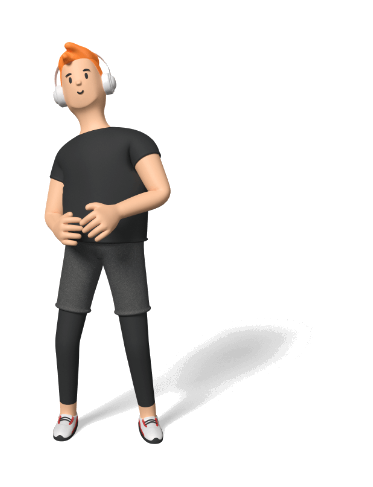}};

\def\ofa{0.4}
\def\ofb{0.9}

\draw [very thick, blue] (1.22, 1.22/2+\ofa) -- plot [smooth, tension=0.6] coordinates { (1.2, 1.2/2+\ofa) (0.9, 0.9/2+\ofb) (0.6, 0.6/2+\ofa) (0.3, 0.3/2+\ofb) (0.0, 0.0/2+\ofa) (-0.3, -0.3/2+\ofb) (-0.6, -0.6/2+\ofa)} -- (-0.62, -0.62/2+\ofa);

\path[fill=sblue,opacity=0.5] (1.22, 1.22/2+\ofa) -- plot [smooth, tension=0.6] coordinates { (1.2, 1.2/2+\ofa) (0.9, 0.9/2+\ofb) (0.6, 0.6/2+\ofa) (0.3, 0.3/2+\ofb) (0.0, 0.0/2+\ofa) (-0.3, -0.3/2+\ofb) (-0.6, -0.6/2+\ofa)} -- (-0.62, -0.62/2+\ofa) -- cycle;

%\draw [very thick, dblue, dotted] (1.82, 1.82/2+\ofa) -- plot [smooth, tension=0.6] coordinates { (1.8, 1.8/2+\ofa) (0.3, 0.3/2+0.7*\ofb) (-1.2, -1.2/2+\ofa)} -- (-1.22, -1.22/2+\ofa);

\path[fill=ssblue,opacity=0.5] (1.82, 1.82/2+\ofa) -- plot [smooth, tension=0.6] coordinates { (1.8, 1.8/2+\ofa) (0.3, 0.3/2+0.7*\ofb) (-1.2, -1.2/2+\ofa)} -- (-1.22, -1.22/2+\ofa) -- cycle;

%\node[inner sep=2pt, anchor=center, opacity=0.7] (oppo) at (0.5, 0.7){\includegraphics[width=1cm]{ov_intersec.png}};

%\node[inner sep=2pt, text width=2.2cm, anchor=north west, fill=white, align=center, opacity=0.9] (cer) at (0.8, 0.2) {\footnotesize Intention without \\ uncertainties};
%\draw [->, >=Stealth, very thick, sblue] ([xshift=-0.6cm] cer.north) -- ([xshift=-1.2cm, yshift=0.4cm] cer.north);

\node[inner sep=2pt, text width=2.2cm, anchor=north west, align=center, fill=vvgray, opacity=0.9] (mmp) at (-3.5, 1.8) {\footnotesize Joint distribution \\ of uncertainties};
\draw [->, >=Stealth, very thick, dyellow] ([xshift=0.6cm] mmp.south) -- (-0.8, 0.7);
\draw [->, >=Stealth, very thick, blue] ([xshift=0.6cm] mmp.south) -- (-0.5, 0.2);

%\node[inner sep=2pt, text width=2.6cm, anchor=north west, align=center, fill=vvgray, opacity=0.9] (ump) at (-3.7, 1.6) {\footnotesize Joint uncertainty with certain intentions};
%\draw [->, >=Stealth, very thick, dorg, dotted] ([xshift=0.6cm] ump.south) -- (-1, 0.6);
%\draw [->, >=Stealth, very thick, dblue, dotted] ([xshift=0.6cm] ump.south) -- (-1, -0.1);

\node[inner sep=2pt, text width=1.8cm, anchor=north west, align=center, fill=vvgray, opacity=0.9] (safe) at (-2.8, -0.2) {\footnotesize Safe trajectory};
\draw [->, >=Stealth, very thick, dgreen] (safe.east) -- (-0.7, -0.5);

\node[inner sep=2pt, text width=2cm, anchor=north west, align=center, fill=vvgray, opacity=0.9] (unsafe) at (0, -0.4) {\footnotesize Unsafe trajectory};
\draw [->, >=Stealth, very thick, dashed, red] (unsafe.west) -- (-0.4, -0.4);

\end{tikzpicture}
\hspace{\fill} 
\caption{An intersection case where an \gls{ev} plans a collision-free left turn, aiming at a safe trajectory for the complex modeling-intention uncertainties of the \gls{ov} and pedestrians. Understanding the intentions of the \gls{ov} and the pedestrians helps the \gls{ev} make safe decisions.}
\label{fig:intersection}
\end{figure}

\IEEEPARstart{W}{hen} an autonomous vehicle interacts with other traffic participants, the awareness of the intentions of these participants strongly impacts the safety and reliability of its decision-making. This has motivated the studies on \textit{intention-aware} motion planning~\cite{park2019planner, wu2023intent, varga2023intention}. Being aware of the intentions of traffic components can notably improve the autonomy and safety of the traffic~\cite{song2022train, moukahal2021vulnerability, hernandez2021ai}.
Fig.~\ref{fig:intersection} shows an intersection case where an \gls{ev} plans a left turn while avoiding collisions with \glspl{ov} and pedestrians from the opposite direction. Understanding the intentions of these participants (e.g., moving fast or slowly) facilitates the design of a safe and flexible policy. Intention-aware motion planning has been formulated as a \gls{pomdp}~\cite{bai2015intention}, where intentions are discrete-valued unobservable variables encoding uncertain policies~\cite{bandyopadhyay2013intention}. 
Decision-making with the awareness of intentions aims at a safe policy for all assumed opponents' behaviors, rendering a challenging planning problem with complex uncertainties, for which planning-based methods, such as reinforcement learning, have been used~\cite{qi2018intent}. However, these methods rely on precise agent models, causing \textit{sim-to-real} issues when models are imprecise~\cite{zhao2020sim}. Moreover, these solutions hardly ensure formal safety guarantees or probable risk restrictions~\cite{li2019formal}. These concerns enforce the need for an intention-aware controller that is correct by design and is robust to modeling errors, which has not been addressed in the literature.

Control of autonomous agents with stochastic uncertainties has been formulated as a stochastic control problem subject to probability constraints~\cite{hoel2023ensemble}. Beyond this, formal safety guarantees have been specified as temporal logic formulas, such as \gls{stl}~\cite{salamati2021data} and \gls{ltl}~\cite{badings2023probabilities}, aiming at a \textit{risk-aware} controller that restricts risk probabilities to certain levels~\cite{engelaar2024risk}. 
A stochastic control problem with formal specifications can be solved using % learning-based control~\cite{saviolo2023active}, 
\gls{mpc}~\cite{farahani2018shrinking} and abstraction-based methods~\cite{haesaert2018temporal}. Belief-space specifications, such as \Gls{poctl}~\cite{jansen2012belief}, \Gls{dtl}~\cite{jones2013distribution}, Gaussian \Gls{dtl}~\cite{leahy2019control}, and probabilistic \Gls{stl}~\cite{sadigh2016safe} have been proposed to convert a stochastic control problem into a tractable deterministic belief-space control problem~\cite{heirung2018stochastic}. Meanwhile, specification decomposition~\cite{leahy2022fast} and splitting~\cite{zhang2023modularized} approaches improve the efficiency of solving a control problem with STL.

However, these existing control methods mainly apply to uncertainties with Gaussian or bounded-support distributions~\cite{lindemann2023safe}. As to be addressed later, intention-aware control requires characterizing complex uncertainties with non-Gaussian or multi-modal distributions, thus remaining an open and challenging problem.
An effective approach to precisely characterize nontrivial probabilistic distributions is stochastic expansion~\cite{eldred2011design}. As a representative method of stochastic expansion, 
\Gls{pce} can efficiently characterize an arbitrary probabilistic model by decomposing it to a linear combination of a finite number of orthogonal bases. It has shown advantages in robust control with non-Gaussian uncertainties~\cite{dai2015distributed}. Stochastic expansion mainly applies to systems with \textit{epistemic} uncertainties~\cite{der2009aleatory} subject to time-invariant distributions. In this sense, stochastic expansion is promising to facilitate the closed-loop intention-aware control for epistemically uncertain systems.

This paper provides the first correct-by-design solution for intention-aware control of uncertain autonomous agents. By formulating unknown intentions of traffic participants as discrete-valued random variables, we derive a stochastic control problem for epistemically uncertain agents with a belief-space specification. In this way, the risk is quantified as the probability of not satisfying certain belief-space specifications, and the control problem is aimed to restrict the risk with a predefined threshold. Then, we use stochastic expansion to convert this challenging stochastic control problem into a tractable deterministic control problem with an \Gls{stl} specification. This allows us to solve a challenging intention-aware control problem using off-the-shelf tools, such as shrinking-horizon \Gls{mpc}~\cite{farahani2018shrinking} based on \Gls{mip}~\cite{kurtz2022mixed}. With assumed intention models, the solved intention-aware controller allows automatic policy adjustment and ensures strict risk limitation. Two autonomous driving cases, namely an overtaking case and an intersection case, have validated the efficacy of this method.

In the remaining part of this paper, Sec.~\ref{sec:frame} formulates the intention-aware control problem, Sec.~\ref{sec:main} defines the mathematical models, Sec.~\ref{sec:solution} presents the mathematical solution to this problem, Sec.~\ref{sec:sim} uses experimental studies to validate the proposed method, Sec.~\ref{sec:disc} discusses the strengthens and weaknesses of our work, and Sec.~\ref{sec:con} concludes the paper.

\textit{Notation:} $\mathbb{R}$($\mathbb{R}^+$), $\mathbb{Z}$($\mathbb{Z}^+$) are sets of (positive) real numbers and integers, respectively. $\mathbb{N}$ denotes natural numbers.

\section{Problem Statement}\label{sec:frame}

% This section states the intention-aware control problem following the exemplary intersection case raised in Sec.~\ref{sec:intro}. In the subsequent sections, we will define the mathematical model for this problem and develop a tractable solution.
 
% \subsection{Intention Awareness}\label{sec:pomdp}

\begin{figure}[htbp]
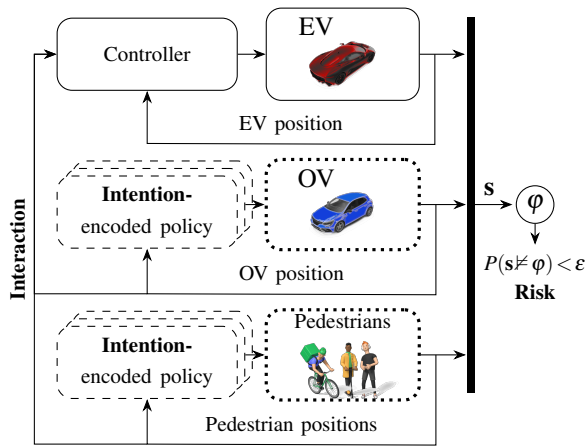

\noindent
\hspace*{\fill} 
\begin{tikzpicture}[scale=1]

\def\sbheight{1.2cm}
\def\sbwidth{1.8cm}
\def\sdev{2cm}

\node[minimum width=0.1cm, inner sep=0pt, minimum height=5cm, fill=black] (mux) at (3cm, 2cm) {};

\node[minimum height=\sbheight,draw, text width=\sbwidth,align=center, rounded corners=2mm] (ev) at (1.3cm, 2*\sdev) {EV~~~~~~~\\ \includegraphics[width=1cm]{ev_intersec.png}};
\node[minimum height=0.8*\sbheight,draw,text width=1.2*\sbwidth,align=center, rounded corners=2mm] (evc) at (-1.3cm, 2*\sdev) {\footnotesize Controller};
\draw [->, >=Stealth] (evc.east) -- (ev.west);
\draw [->, >=Stealth] (ev.east) -- ([xshift=0.6cm] ev.east);
\draw [->, >=Stealth] ([xshift=0.2cm] ev.east) -- ([xshift=0.2cm, yshift=-1.2cm] ev.east) -- node[pos=0.5, anchor=south]{\footnotesize EV position} ([yshift=-0.7cm] evc.south) -- (evc.south);

\node[minimum height=\sbheight,draw, very thick, dotted, text width=\sbwidth,align=center, rounded corners=2mm] (ov) at (1.3cm, \sdev) {OV~~~~~~~\\ \includegraphics[width=1cm]{ov_intersec.png}};

\node[minimum height=0.8*\sbheight,draw,text width=1.2*\sbwidth,align=center, fill=white, dashed, rounded corners=2mm] (ovc1) at (-1.1cm, \sdev+0.1cm) {\footnotesize \textbf{Intention-} \\ encoded policy};
\node[minimum height=0.8*\sbheight,draw,text width=1.2*\sbwidth,align=center, fill=white, dashed, rounded corners=2mm] (ovc2) at ([xshift=-0.1cm, yshift=-0.1cm] ovc1) {\footnotesize \textbf{Intention-} \\ encoded policy};
\node[minimum height=0.8*\sbheight,draw,text width=1.2*\sbwidth,align=center, fill=white, dashed, rounded corners=2mm] (ovc3) at ([xshift=-0.1cm, yshift=-0.1cm] ovc2) {\footnotesize \textbf{Intention-} \\ encoded policy};

\draw [->, >=Stealth] (ovc2.east) -- (ov.west);
\draw [->, >=Stealth] (ov.east) -- ([xshift=0.6cm] ov.east);
\draw [->, >=Stealth] ([xshift=0.2cm] ov.east) -- ([xshift=0.2cm, yshift=-1.2cm] ov.east) -- node[pos=0.5, anchor=south]{\footnotesize OV position} ([yshift=-0.6cm] ovc3.south) -- (ovc3.south);
\draw [->, >=Stealth] ([yshift=-0.6cm] ovc3.south) -- ([xshift=-1.5cm, yshift=-0.6cm] ovc3.south) -- ([xshift=-1.5cm] evc.center) -- (evc.west);

\node[minimum height=\sbheight,draw, very thick, dotted,text width=\sbwidth,align=center, rounded corners=2mm] (pd) at (1.3cm, 0cm) {};
\node[minimum height=0.8*\sbheight,draw,text width=1.2*\sbwidth,align=center, dashed, rounded corners=2mm] (pdc1) at (-1.1cm, 0.1cm) {\footnotesize \textbf{Intention}};
\node[minimum height=0.8*\sbheight,draw,text width=1.2*\sbwidth,align=center, fill=white, dashed, rounded corners=2mm] (pdc2) at ([xshift=-0.1cm, yshift=-0.1cm] pdc1) {\footnotesize \textbf{Intention}};
\node[minimum height=0.8*\sbheight,draw,text width=1.2*\sbwidth,align=center, fill=white, dashed, rounded corners=2mm] (pdc3) at ([xshift=-0.1cm, yshift=-0.1cm] pdc2) {\footnotesize \textbf{Intention-}\\ encoded policy};
\draw [->, >=Stealth] (pdc2.east) -- (pd.west);
\draw [->, >=Stealth] (pd.east) -- ([xshift=0.6cm] pd.east);
\draw [->, >=Stealth] ([xshift=0.2cm] pd.east) -- ([xshift=0.2cm, yshift=-1.2cm] pd.east) -- node[pos=0.5, anchor=south]{\footnotesize Pedestrian positions} ([yshift=-0.6cm] pdc3.south) -- (pdc3.south);
\draw [->, >=Stealth] ([yshift=-0.6cm] pdc3.south) -- ([xshift=-1.5cm, yshift=-0.6cm] pdc3.south) -- node[anchor=south, rotate=90] {\footnotesize \textbf{Interaction}} ([xshift=-1.5cm] evc.center) -- (evc.west);

\node[inner sep=2pt, anchor=center] (ped1) at ([xshift=0.5cm, yshift=-0.2cm] pd.center) {\includegraphics[width=0.7cm]{ped1.png}};
\node[inner sep=2pt, anchor=center] (ped2) at ([xshift=-0.3cm, yshift=0.0cm] ped1) {\includegraphics[width=0.8cm]{ped2.png}};
\node[inner sep=2pt, anchor=center] (ped3) at ([xshift=-0.7cm, yshift=0.0cm] ped1) {\includegraphics[width=1cm]{ped3.png}};

\node[inner sep=2pt, anchor=north] () at (pd.north) {\footnotesize Pedestrians};

\draw [->, >=Stealth] (mux.east) -- node[pos=0,anchor=south west,align=left]{$\sseq$} ([xshift=0.5cm] mux.east);

\node[circle,inner sep=0pt,minimum size=0.5cm,draw] (phi) at ([xshift=0.8cm] mux.east) {};
\node[] at (phi) {$\psi$};

\node[minimum height=0.5cm, minimum width=0.5cm, align=center] (risk) at ([yshift=-0.8cm] phi) {\footnotesize $P(\sseq \!\nvDash\! \psi) \!<\! \varepsilon$};
\node[] at ([yshift=-0.1cm] risk.south) {\footnotesize \textbf{Risk}};
\draw [->, >=Stealth] (phi.south) -- (risk.north);

\end{tikzpicture}
\hspace{\fill} 
\caption{The intention-aware control framework.}
\label{fig:structure}
\end{figure}

The intersection case shown in Fig.~\ref{fig:intersection} serves as a typical autonomous driving scenario where an \gls{ev} needs to plan a safe trajectory incorporating the interaction with the other traffic participants, such as \glspl{ov} and pedestrians. In this case, the intention-aware control framework is illustrated in Fig.~\ref{fig:structure} which will be interpreted in detail as follows.

\subsection{Description of Intention-Aware Control}\label{sec:uitp}

As an interactive participant, an \gls{ov} or a pedestrian seeing the \gls{ev} approaching might have various intents that encode different driving or moving policies~\cite{dang2023identifying}, requiring the \gls{ev} to take different strategies accordingly. For example, an antagonistic \gls{ov} tends to \textit{speed up} to show a strong sense of priority. On the contrary, a cooperative \gls{ov} may \textit{slow down} to give out the priority. There might also be an ignorant \gls{ov} that remains a \textit{constant speed}, regardless of the reaction to the \gls{ev}. Similarly, younger pedestrians tend to move faster than the elders. A participant's intention may be random, meaning that which intent will be true is not known. In Fig.~\ref{fig:structure}, we use dashed blocks to denote the policies of the traffic participants encoded by uncertain intentions. Moreover, the behaviors of the traffic participants may be dominated by uncertain parameters unknown to the \gls{ev}, such as the sizes of vehicles or the step sizes of pedestrians. Thus, in Fig.~\ref{fig:structure}, we use dotted blocks to represent these participants. Due to uncertain intentions and parameters, the behaviors of these traffic participants are random. In this sense, we are interested in designing a controller for the \gls{ev} with mitigated risk, or minimal likelihood of collisions, by incorporating all possible behaviors of other traffic participants dominated by their unknown intentions. We refer to such a controller as an \textit{intention-aware} controller.

\subsection{Problem Formulation}\label{sec:iacwf}

An \textit{intention-aware} controller allows an autonomous vehicle to interact with other interactive traffic participants that proactively react to itself, forming a foundation of \textit{interaction-aware} planning~\cite{lefkopoulos2020interaction}. Let us see the joint positions of all traffic participants as a random signal denoted as $\sseq:=s_0s_1\cdots$ and describe the driving task using an \gls{stl} specification $\psi$ (see Appx.-A for \gls{stl}). We aim to solve a controller to ensure a restricted probability $P(\sseq \vDash \psi) > 1 - \varepsilon$, meaning that the intersection task should be accomplished with a minimal probability $1-\varepsilon$, where $\varepsilon \in (0,1)$ is a predefined risk level. This renders a \textit{risk-aware} policy by imposing a strict limitation on the risk probability $P(\sseq \nvDash \psi) \!\leq\! \varepsilon$, as shown in  Fig.~\ref{fig:structure}. Then, the intention-aware control problem is formulated as follows.

\begin{mdframed}[linewidth=1pt, 
        backgroundcolor=lightgray!20  
        ]
\begin{problem}[Intention-Aware Control]\label{pb:int}
Consider an autonomous (ego) vehicle and multiple traffic participants with uncertain intentions and parameters. Solve a controller for the ego vehicle, such that the joint positions of all participants, denoted as $\sseq:=s_0 s_1 \cdots$, satisfies a predefined \gls{stl} specification $\psi$ at a restricted level $P(\sseq \vDash \psi) > 1 - \varepsilon$, with $\varepsilon \!\in\!(0,1)$.
\qed
\end{problem}\end{mdframed}

In Problem~\ref{pb:int}, by investigating the risk measure $P(\sseq \vDash \psi)$, we ignore the influence of historical observations and assume that the uncertainty characteristics of all traffic participants denoted by $\sseq$ are known. Although not always realistic in practical scenarios, this assumption simplifies the problem and allows us to focus on characterizing nontrivial uncertainty distributions. However, such a simplification still leads to a critical challenge of characterizing the complex joint uncertainty brought up by random intentions and parameters of the traffic participants. In the following section, we will give a mathematical solution to Problem~\ref{pb:int} by leveraging stochastic expansion methods. We will address extending our method by releasing this assumption in Sec.~\ref{sec:disc}.

\section{Mathematical Models}\label{sec:main}

This section defines the mathematical models of the traffic participants in a general traffic scenario with epistemic uncertainty and belief-space specifications. For clarity, intuitive explanations will follow to interpret how these models are connected with practical traffic scenarios.

\subsection{Modeling Traffic Participants}\label{sec:systm}

From the ego perspective, uncertainty mainly originates from the lack of knowledge of its opponents. In this sense, we consider a general traffic system containing a deterministic \gls{ev} and $M\!\in\!\mathbb{Z}^+$ uncertain opponent participants. Without losing generality, we assume all opponent participants are vehicles (\glspl{ov}). The \gls{ev} is described as the following deterministic dynamic model, as commonly used in the literature~\cite{brudigam2021stochastic},
\begin{equation}\label{eq:ego}
\mathrm{EV}: \left\{ \begin{array}{l}
x_{k+1} = A x_k + B u_k + d_k,  \\
~~~y_k = C x_k,
\end{array} \right. 
\end{equation}
where $x_k\!\in \! \mathbb{R}^n$, $u_k\!\in\!  \mathbb{R}^{m}$, $d_k \!\in\! \mathbb{R}^d$, and $y_k \!\in \! \mathbb{R}^l$ are the state, the control input, the external input, and the measurement output of the \gls{ev}, respectively, at time $k \in \mathbb{N}$, and $A \in \mathbb{R}^{n \times n}$, $B \in \mathbb{R}^{n \times m}$, $C \in \mathbb{R}^{l \times n}$ are constant matrices, where $n,m,l \in \mathbb{Z}^+$. Also, $d_k$ for $k\!\in\!\mathbb{N}$ is a bounded external input denoting the deterministic disturbance or modeling errors of the \gls{ev}. A commonly used bicycle model can be linearized as Eq.~\eqref{eq:ego}, as addressed in Appx.-B.

%Consider $M \in \mathbb{N}^+$ \glspl{ov}, e
Meanwhile, each \gls{ov} is described as a stochastic linear dynamic model, as commonly used in the literature~\cite{fagiano2012nonlinear},
\begin{equation}\label{eq:oppo}
\mathrm{OV}_j: \left\{ \begin{array}{l}
z_{k+1}^{(j)} = F^{(j)}(\theta^{(j)}) z_{k}^{(j)} + G^{(j)}(\theta^{(j)}) v_{k}^{(j)}, \\
w_{k}^{(j)} = H^{(j)}(\theta^{(j)}) z_{k}^{(j)},
\end{array} \right.
\end{equation}
where $z_{k}^{(j)}$, $v_{k}^{(j)}$, and $w_{k}^{(j)}$ are state, control input, and output of the $j$-th \gls{ov} with certain dimensions, $F^{(j)}$, $G^{(j)}$, and $H^{(j)}$ are parametric matrices dependent on a random parameter $\theta^{(j)}$ subject to a multi-dimensional joint \gls{pdf} $p_{\theta^{(j)}}$. All vectors and matrices are of proper dimensions. We assume the elements of $\theta^{(j)}$ are independent and identically distributed (i.i.d) random variables with finite second-order moments.  Here, the random parameter $\theta^{(j)}$ represents the guess of the ego about the physical features of the \glspl{ov}, such as length or friction. Besides, Eq.~\eqref{eq:oppo} allows a heterogeneous setting, where the \glspl{ov} may be of different dimensions and parametric matrices. A nonlinear bicycle model can be represented using the general uncertain model in Eq.~\eqref{eq:oppo} after linearization, as addressed in Appx.-C. 
In this paper, we also use an \gls{ov} model defined in~\eqref{eq:oppo} to represent the behavior of a group of pedestrians~\cite{ahn2021safe}. 

The overall model of all opponent traffic participants can be represented as the following compact form,
\begin{equation}\label{eq:linear_dy_compact_op}
\mathrm{OVs}: \left\{ \begin{array}{l}
z_{k+1} = F(\theta) z_{k} + G(\theta)v_k, \\
~~w_k = H(\theta) z_{k},
\end{array} \right. 
\end{equation}
where $z_k \in \mathbb{R}^p$, $v_k \in \mathbb{R}^q$, $w_k \in \mathbb{R}^r$, and $\theta \in \mathbb{R}^t$ are vectors with the joint state, input, output, and uncertain parameters of all \glspl{ov}, respectively, and $F \!\in\! \mathbb{R}^{p \times p}$, $G \!\in\! \mathbb{R}^{p \times q}$, and $H \!\in\! \mathbb{R}^{r \times q}$ are augmented matrices containing the parametric matrices of the \glspl{ov}, with $p,q,r,t\in\mathbb{Z}^+$ being proper dimensions.

\subsection{Random Intentions with Epistemic Uncertainty}\label{sec:sys_int}

For each \gls{ov}-$j$, we use a discrete-valued random variable $\iota^{(j)}$ sampled from a finite intent set $I^{(j)}$ with a probabilistic distribution $p_{\iota^{(j)}}$ to represent its unknown intention. Based on this, we assume that each \gls{ov} is controlled by the following intention-encoded policy,
\begin{equation}\label{eq:linear_policy}
\textstyle v_{k}^{(j)}:= K^{(j)}(\iota^{(j)})w_{k}^{(j)} + L^{(j)}(\iota^{(j)})\tau_{k}^{(j)}, 
\end{equation}
where $\tau_{k}^{(j)}$ is a predefined feedforward control input with proper dimensions, and $K^{(j)}$ and $L^{(j)}$ are the feedback and the feedforward control gain matrices of proper dimensions, encoded by the intention $\iota^{(j)}$, respectively. Such an intention-encoded policy reflects how the ego assumes the driving style of an \gls{ov}, such as \textit{antagonistic}, \textit{ignorant}, and \textit{cooperative}. Understanding all possible driving styles of the \glspl{ov} helps the \gls{ev} generate a reasonable control policy. Moreover, Eq.~\eqref{eq:linear_policy} is defined to cover the most common control modes in practice, namely feedback control, feedforward control, and interaction with other \glspl{ov}, for representativeness.

Substituting Eq.~\eqref{eq:linear_policy} to Eq.~\eqref{eq:oppo} and rewriting it in a compact form as~Eq.~\eqref{eq:linear_dy_compact_op}, we obtain the following closed-loop dynamic model for the \glspl{ov},
\begin{equation}\label{eq:linear_dy}
\textstyle z_{k+1} = F(\theta)z_k + G(\theta) K(\iota) H(\theta) z_{k} + G(\theta) L(\iota) \tau_{k},
\end{equation}
where $\iota$ and $\tau_k$ are the joint intention and joint feedforward control inputs of the \glspl{ov}, respectively, and $K$ and $L$ are control gain matrices encoded by the joint intention $\iota$. Then, Eq.~\eqref{eq:linear_dy} can be compactly written as
\begin{equation}\label{eq:linear_dy_compact}
\mathrm{OVs\,(closed~loop)}: \left\{ \begin{array}{l}
z_{k+1} = \mathbf{F}(\eta) z_{k} + \mathbf{G}(\eta) \tau_k, \\
~~w_k = \mathbf{H}(\eta) z_{k},
\end{array} \right. 
\end{equation}
where $\eta:=[\,\theta^{\top}\,\iota^{\top}]^{\top}$ denotes the joint uncertainties of all \glspl{ov}, $\mathbf{F}$ and $\mathbf{G}$ are joint matrices derived from Eq.~\eqref{eq:linear_dy}, and $\mathbf{H}(\eta):=H(\theta)$. In this sense, the closed-loop model in Eq.~\eqref{eq:linear_dy_compact} characterizes all possible behaviors of the \glspl{ov}. Due to the continuous nature of $\theta$ and the discrete nature of $\iota$, the random variable $\eta$ is usually subject to a complex distribution which is difficult to characterize.

%\subsection{Epistemic Uncertainty}

With an arbitrary initial condition $x_0 \in \mathbb{R}^n$, the \gls{ev} model in \eqref{eq:ego} generates a deterministic output trajectory $\yseq:=y_0y_1y_2\cdots$. In the meantime, with the influence of the random variable $\eta$, the \gls{ov} model in \eqref{eq:linear_dy_compact} generates a stochastic output trajectory $\wseq:=w_0 w_1 w_2\cdots$. Due to the complex distribution of $\eta$, the trajectories $\wseq$ may follow a complex multi-modal join distribution, as illustrated in Fig.~\ref{fig:intersection}. Quantifying such complex uncertainty poses a challenge to solving the intention-aware control problem. Fortunately, such uncertainty has an \textit{epistemic} nature since the distribution of $\eta$ is time-invariant and can be learned with sufficient data~\cite{der2009aleatory}. On the contrary is \textit{aleatory} uncertainty that can never be precisely characterized by a time-invariant probabilistic distribution, such as random noise~\cite{mckeand2021stochastic}. Epistemic uncertainty can be precisely and tractably characterized using stochastic expansion methods like \gls{pce}~\cite{eldred2011design}. In this sense, \gls{pce} shows great potential in solving the intention-aware control problem that requires precise quantification of epistemic uncertainty. To simplify the situation and focus on its control aspect, we assume that the distribution of $\eta$ is known and is not affected by historical observations. This means that the \gls{ev} knows the stochastic characteristics of the dynamics and intentions of all \glspl{ov}, even though it is unaware of their exact values. We will address how the solution can be extended to incorporate historical observations in Sec.~\ref{sec:disc}.

% Even the exact values of the random parameters and intentions are not known, their distributions can be obtained using empirical or inference methods~\cite{xing2019driver, gangopadhyay2019identification}. 

% Stochastic expansion methods, such as \gls{pce}, have provided powerful tools to precisely characterize the propagation of complex distributions~\cite{eldred2011design}. The characterization has proved to be tractable for time-invariant distributions. 

\subsection{Quantifying Risk Using Probabilistic Constraints}\label{sec:risk}

Practical driving tasks can be represented as physical constraints for the traffic participants. Violating these constraints leads to the failure of the tasks. For participants with stochastic uncertainties as defined in Sec.~\ref{sec:systm}, a typical physical constraint has the following probabilistic form~\cite{hoel2023ensemble},
\begin{equation}\label{eq:constr}
\textstyle P\!\left( \gamma^{\!\top\!} w_k \!\leq\! \beta - \alpha^{\top\!} y_k \right) \geq 1-\varepsilon,~k \in \mathbb{N},
\end{equation}
where $y_k \in \mathbb{R}^l$, $w_k \in \mathbb{R}^r$ are outputs of the \gls{ev} and the \glspl{ov} given by Eq.~\eqref{eq:ego} and Eq.~\eqref{eq:linear_dy_compact}, respectively, $\alpha$, $\beta$, $\gamma$ are constant coefficients with proper dimensions, and $\varepsilon \!\in\![\,0,\,1\,]$ is a risk restriction scalar. Eq.~\eqref{eq:moment_constr} prescribes that the probability of the \gls{ev} and the \gls{ov} satisfying a desired condition $\gamma^{\!\top\!} w_k \!\leq\! \beta - \alpha^{\top\!} y_k$ should be no smaller than a predefined threshold $1-\varepsilon$. Based on this, we can use the complimentary probability $P\!\left( \gamma^{\!\top\!} w_k \!>\! \beta - \alpha^{\top\!} y_k \right)$ to quantify the risk of violating $\gamma^{\!\top\!} w_k \!\leq\! \beta - \alpha^{\top\!} y_k$. In this sense, Eq.~\eqref{eq:moment_constr} renders a risk restriction that the complimentary probability $P\!\left( \gamma^{\!\top\!} w_k \!>\! \beta - \alpha^{\top\!} y_k \right)$ should be no bigger than a predefined threshold $\varepsilon$. 

It is typically challenging to handle probabilistic constraints as Eq.~\eqref{eq:constr} directly. A common way is to convert it to a deterministic constraint defined in a certain probability space~\cite{hoel2023ensemble}. If $w_k$ is Gaussian for all $k \in \mathbb{N}$, according to~\cite{calafiore2006distributionally}, the probabilistic constraint Eq.~\eqref{eq:constr} is equivalent to
\begin{equation}\label{eq:moment_constr}
\alpha^{\top} y_k + \gamma^{\!\top}\mu_{w_k} + \kappa_{\varepsilon} \sqrt{\gamma^{\!\top} \Sigma_{w_k} \gamma} -\beta \leq 0,
\end{equation}
where $\mu_{w_k}\!\in\!\mathbb{R}^r$ and $\Sigma_{w_k}\!\in\!\mathbb{R}^{r\times r}$ are the expected value and the covariance of $w_k$,  respectively, and $\kappa_{\varepsilon} \!=\! \Psi^{-1}(1-\varepsilon)$ is a scalar with $\Psi$ being the cumulative probability function of the normal distribution. However, the equivalence between Eq.~\eqref{eq:constr} and Eq.~\eqref{eq:moment_constr} does not hold if $w_k$ is non-Gaussian, for which conservativeness is inevitably introduced. Consider that $w_k$ has finite statistic moments of all orders. Let $\mathcal{D}(\mu_{w_k}, \Sigma_{w_k})$ be the set of all random variables that have the same expected value and covariance as $w_k$. Then, the following constraint,
\begin{equation}\label{eq:moment_constr_cons}
\mathrm{inf}_{\omega\in\mathcal{D}(\mu_{w_k},\Sigma_{w_k})} P( \gamma^{\!\top\!} \omega \!\leq\! \beta - \alpha^{\top\!} y_k ) \geq 1- \varepsilon,~k \in \mathbb{N},
\end{equation}
is equivalent to Eq.~\eqref{eq:moment_constr} but with $\kappa_{\epsilon} = \sqrt{(1-\epsilon)/\epsilon}$~\cite{calafiore2006distributionally}. Note that Eq.~\eqref{eq:moment_constr_cons} is conservative to Eq.~\eqref{eq:constr} due to the $\mathrm{inf}$ operator. 

Let $\mathcal{B}^r := \mathbb{R}^r \times \mathbf{S}^r$ be a belief space of $r$-dimensional random variables with known first- and second-order moments, where $\mathbf{S}^r \subset \mathbb{R}^{r \times r}$ is the set of all positive semi-definite matrices, such that $(\mu_{w_k}, \Sigma_{w_k}) \in \mathcal{B}^r$, $\forall \, k \in \mathbb{N}$. We can address that a probability constraint as Eq.~\eqref{eq:constr} can be transformed into a sufficient deterministic constraint as Eq.~\eqref{eq:moment_constr} defined on the belief space $\mathcal{B}^r$. %In this way, we build a connection between probabilistic constraints and belief-space inequalities, that is, any probabilistic constraint in the form of Eq.~\eqref{eq:constr} corresponds to a sound belief-space inequality as Eq.~\eqref{eq:moment_constr}.

\subsection{Describing Tasks Using Belief-Space Specifications}\label{sec:dstl}

The probability constraint as Eq.~\eqref{eq:constr} can only specify the states of the participants at a certain time $k \in \mathbb{N}$. In practice, a driving task should be described using multiple constraints for different times which can be specified using temporal logic specifications~\cite{salamati2021data}. 
In Sec.~\ref{sec:risk}, we have demonstrated how a probability constraint can be converted to a sufficient deterministic belief-space constraint. Now, we extend the belief-space inequalities as Eq.~\eqref{eq:moment_constr} into belief-space formal specifications as follows.

\begin{definition}[\gls{dstl}]\label{def:dtl}
For random signals $\yseq:=y_0 y_1 \cdots$ generated by Eq.~\eqref{eq:ego} and $\wseq:= w_0 w_1 \cdots$ generated by Eq.~\eqref{eq:linear_dy_compact}, a \gls{dstl} formula $\varphi$ defined over the joint signal 
$(\yseq, \wseq)\!:=\!(y_0, w_0)(y_1, w_1)\cdots$ has the following inductively defined syntax,
\begin{equation*}
\varphi \,:= \top \,|\, (\nu, \varepsilon) \,|\, \lnot \varphi \,|\, \varphi_1 \!\wedge\! \varphi_2 \,|\, \varphi_1 \mathsf{U}_{[\,a,\,b\,]} \varphi_2,
\end{equation*}
where $\varphi$, $\varphi_1$ and $\varphi_2$ are \Gls{dstl} formulas, $(\nu, \varepsilon)$ is a belief-space predicate associated with a mapping $g_{\varepsilon}: \mathbb{R}^l \times \mathcal{B}^r \rightarrow \mathbb{R}$ via $\displaystyle (\nu, \varepsilon) \!:=\! \left\{ \begin{array}{ll}
\top, &g_{\varepsilon}(y_k, (\mu_{w_k}, \Sigma_{w_k})) \leq 0 \\
\lnot \top, & g_{\varepsilon}(y_k, (\mu_{w_k}, \Sigma_{w_k})) > 0
\end{array} \right.$, for a given time $k\!\in\!\mathbb{N}$, where $g_{\varepsilon}$ is defined as the left hand of the inequality Eq.~\eqref{eq:moment_constr}, and $\mathsf{U}_{[\,a,\,b\,]}$ is the \textit{until} operator bounded with time interval $[\,a,\,b\,]$, where $a,b \in \mathbb{N}$ and $a \leq b$. We define the \textit{disjunction}, \textit{eventually}, and \textit{always} operators as $\varphi_1 \!\vee\! \varphi_2 \!:=\! \lnot \left(\lnot \varphi_1 \wedge \lnot \varphi_2 \right)$, 
$\mathsf{F}_{[\,a,\,b\,]} \varphi \!:=\! \top \mathsf{U}_{[\,a,\,b\,]} \varphi$,
$\mathsf{G}_{[\,a,\,b\,]} \varphi\!:=\! \lnot \left( \top \mathsf{U}_{[\,a,\,b\,]} \lnot \varphi \right)$. 
The semantics of the \gls{dstl} $((\yseq, \wseq),\,k) \vDash \varphi$ are recursively defined as follows,
{\small
\begin{equation*}
\begin{array}{l}
((\yseq, \wseq),\,k) \vDash (\nu, \varepsilon)  \leftrightarrow  g_{\varepsilon}(y_k, (\mu_{w_k}, \Sigma_{w_k})) \leq 0, \\
((\yseq, \wseq),\,k) \vDash \lnot \varphi  \leftrightarrow  \lnot (((\yseq, \wseq),\,k) \vDash \varphi), \\
((\yseq, \wseq),\,k) \vDash \varphi_1 \wedge \varphi_2  \leftrightarrow  ((\yseq, \wseq), k) \vDash \varphi_1 \wedge ((\yseq, \wseq),\,k) \vDash \varphi_2, \\
((\yseq, \wseq),\,k) \vDash \varphi_1 \vee \varphi_2  \leftrightarrow  ((\yseq, \wseq), k) \vDash \varphi_1 \vee ((\yseq, \wseq),k) \vDash \varphi_2, \\
((\yseq, \wseq),\,k) \vDash \mathsf{F}_{[\,a,\,b\,]} \varphi  \leftrightarrow  \exists \, k' \!\in\! [\,k\!+\!a,\,k\!+\!b\,], ((\yseq, \wseq),\,k') \vDash \varphi, \\
((\yseq, \wseq),\,k) \vDash \mathsf{G}_{[\,a,\,b\,]} \varphi  \leftrightarrow  \forall \, k' \!\in\! [\,k\!+\!a,\,k\!+\!b\,], ((\yseq, \wseq),\,k') \vDash \varphi, \\
((\yseq, \wseq),\,k) \vDash \varphi_1 \mathsf{U}_{[\,a,\,b\,]} \varphi_2  \leftrightarrow  \exists \, k' \in [\,k+a,\,k+b\,], \\
~~\mathrm{s.t.} ((\yseq, \wseq),\,k') \vDash \varphi_2,~ \mathrm{and} ~ \forall \, k'' \in [\,k, \,k'\,],((\yseq, \wseq),\,k'') \vDash \varphi_1.
\end{array}
\end{equation*}}
For $k=0$, the symbol $k$ is omitted, rendering  $(\yseq, \wseq) \vDash \varphi $.
\qed
\end{definition}

DSTL can be understood straightforwardly. Each \gls{dstl} predicate renders a belief-space inequality as Eq.~\eqref{eq:moment_constr_cons} that is sufficient for a probabilistic constraint like Eq.~\eqref{eq:constr}. Then, a DSTL formula can be seen as a set of belief-space inequalities for multiple probability constraints specified for different times $k$. In this sense, DSTL can conveniently specify driving tasks for traffic participants with stochastic uncertainties. Failing to satisfy the DSTL specification indicates the failure of the entire driving task. The risk of task failure can be quantified as the probability that a DSTL formula is not satisfied, namely $P((\yseq, \wseq) \!\nvDash\! \varphi)$ which is an alternative to the original STL-based risk $P(\sseq \!\nvDash\! \psi)$ as illustrated in Fig.~\ref{fig:structure}.

\Gls{dstl} can be recognized as a generalization of \gls{stl} (See Appx.-A) to a belief space or an extension of \gls{dtl}~\cite{jones2013distribution} or Gaussian \gls{dtl}~\cite{leahy2019control} to real-valued signals. Its main advantage over the conventional STL is to generate deterministic belief-space constraints directly without rendering probability constraints. Note that the belief space considered in this paper does not depend on historical observations, implying that the risk $P((\yseq, \wseq) \!\nvDash\! \varphi)$ considered in this paper does not incorporate historical observations. We will discuss potential solutions for possible extensions in sec.~\ref{sec:disc}.

\section{Tractable Control Solution Based on PCE}\label{sec:solution}

With the mathematical models defined in Sec.~\ref{sec:main}, Problem~\ref{pb:int} can be reformulated as the following formal control problem specified by a DSTL formula and solved using the following \textit{shrinking-horizon} \Gls{mpc}-based approach~\cite{farahani2018shrinking} within a predefined control horizon $\mathcal{H}\!:=\!\{0,1,\cdots,N-1\}$, $N \in \mathbb{Z}^+$. 

\begin{mdframed}[linewidth=1pt, 
backgroundcolor=lightgray!20  
        ]
    \begin{problem}[Formal Intention-Aware Control]\label{pb:intent}
For the mathematic models defined in Sec.~\ref{sec:main}, solve a history- and intention-dependent controller $\pi:\mathbf{Y}^+ \times  \mathcal{P}\!\rightarrow\! \mathbf{U}$ for the \gls{ev}, where $\mathbf{Y}^+$ is the set of all output trajectories of the \gls{ev}, $\mathcal{P}$ is the space of all possible distributions of the intentions of the \glspl{ov}, and $\mathbf{U}$ is the feasible control set of the \gls{ev}, such that the output trajectories $\yseq$ and $\wseq$ satisfy a predefined \gls{dstl} formula, i.e., $(\yseq, \wseq)\vDash \varphi$.
\qed
\end{problem}\end{mdframed}

\begin{mdframed}[linewidth=1pt, 
backgroundcolor=lightgray!20  
        ]
        \textbf{Shrinking-Horizon MPC-based Solution:}
For any time $k \in \mathcal{N}$ with historical measurements $y_0y_1\cdots y_k \!\in\! \mathbf{Y}^+$, solve the following stochastic optimization problem,
\begin{subequations}\label{eq:opti_sto}
\begin{alignat}{6}
&\textstyle \min_{\tilde{\useq}_k}  \sum_{i=0}^{N-k-1} \tilde{u}_{i|k}^{\top}R \tilde{u}_{i|k}
\label{eq:min} \\
\mathrm{s.t.}~&\tilde{x}_{i+1|k} \!= A \tilde{x}_{i|k} + B \tilde{u}_{i|k},\\
&\tilde{u}_{i|k} \in \mathbf{U},~\forall \, i\!\in\! \{0,1,\cdots,N\!-\!k\!-\!1\},\\
&\tilde{y}_{i+k|k} \!= C \tilde{x}_{i|k},~\forall \, i\in \{0,1,\cdots,N-k\}, \\
& \tilde{y}_{r|k} = y_r,~\forall \, r\in \{0,1,\cdots,k\},\label{eq:hist} \\
& z_{i+1} \!= \mathbf{F}(\eta) z_i + \mathbf{G}(\eta) \tau_i,~\forall \, i\!\in\! \mathcal{H}, \\
& w_i  \!= \mathbf{H}(\eta) z_i,~\forall \, i\!\in\! \mathcal{H} \cup \{N\}, \\
& (\tilde{\yseq}_k, \wseq) \vDash \varphi, \label{eq:spec}
\end{alignat}
\end{subequations}
where $\tilde{\xseq}_k:=\tilde{x}_{0|k} \cdots \tilde{x}_{N-k|k}$, $\tilde{\useq}_k:=\tilde{u}_{0|k}\cdots \tilde{u}_{N-k-1|k}$, and $\tilde{\yseq}_k:=\tilde{y}_{0|k} \cdots \tilde{y}_{N|k}$ are decision variables for the state, the input, and the output at time $k$, $R \in \mathbb{R}^{m \times m}$ is a cost matrix, $\mathbf{U} \subseteq \mathbb{R}^m$ is a control constraint set, and $\wseq:=w_0 w_1 \cdots w_N$ is the measurement trajectory of the \glspl{ov} generated by Eq.~\eqref{eq:oppo} with an intention-dependent policy in \eqref{eq:linear_policy}. Having solved $\tilde{\useq}_k$, apply its first element $\tilde{u}_{0|k}$ to the \gls{ev} model in Eq. \eqref{eq:ego} as the control input. 
\end{mdframed}
\begin{remark}[Robustness]
The control scheme above is closed-loop since the control input $\tilde{u}_{0|k}$ at time $k\in \mathcal{N}$ depends on the historical measurements $y_0 y_1 \cdots y_k$. Thus, the controller is robust to the unknown disturbance $d_k$ in Eq.~\eqref{eq:ego}.
\end{remark}

The solution above is challenging due to the complex joint uncertainty brought up by $\eta$. Now, we provide an efficient control solution facilitated by \gls{pce}. As a typical stochastic expansion approach, \gls{pce} allows the stochastic parameters $\mathbf{F}$, $\mathbf{G}$, and $\mathbf{H}$ in Eq.~\eqref{eq:linear_dy_compact} to be approximated as $\mathbf{F} \!=\! \sum_{i=0}^{L-1} \mathbf{\hat{F}}_i \Phi_i$, $\mathbf{G} \!=\! \sum_{i=0}^{L-1} \mathbf{\hat{G}}_i \Phi_i$, and $\mathbf{H} \!=\! \sum_{i=0}^{L-1} \mathbf{\hat{H}}_i \Phi_i$, where $L \in \mathbb{Z}^+$ is a predefined finite expansion order, $\pmb{\Phi}\!=\!\left\{\Phi_0~\Phi_1~\cdots~\Phi_{L-1} \right\}$ is a series of orthogonal polynomial bases with $\Phi_0:=1$, and $\mathbf{\hat{F}}_i \!\in\!\mathbb{R}^{p \times p}$, $\mathbf{\hat{G}}_i\!\in\!\mathbb{R}^{p \times q}$, and $\mathbf{\hat{H}}_i\!\in\!\mathbb{R}^{r \times p}$, for $i \in \{0,1,\cdots, L-1\}$, are PCE coefficients which can be calculated using off-the-shelf tools~\cite{feinberg2018multivariate}. The approximation precision is guaranteed in a Hilbert sense~\cite{feinberg2018multivariate}. Similar expansion can also be performed to $z_k$ and $w_k$, with $z_k \!= \!\sum_{i=0}^{L-1} \hat{z}_{k,i} \Phi_i,$ and $w_k \!= \!\sum_{i=0}^{L-1} \hat{w}_{k,i} \Phi_i$, for $k \in \mathcal{N}$, 
where $\hat{z}_{k,i} \in \mathbb{R}^{p}$ and $\hat{w}_{k,i} \in \mathbb{R}^{r}$ are expansion coefficients of $z_k$ and $w_k$, respectively. Applying the above expansions to Eq.~\eqref{eq:linear_dy_compact}, we obtain the following PCE-based dynamic model,
\begin{equation}\label{eq:deter_dy}
\mathrm{OVs\,(PCE~model)}: \left\{ \begin{array}{l}
\hat{z}_{k+1} \!= \mathbf{\hat{F}} \hat{z}_k + \mathbf{\hat{G}} \tau_k,\\
~~\hat{w}_k  \!= \mathbf{\hat{H}} \hat{z}_k,
\end{array} \right. 
\end{equation}
where $\hat{z}_k \!=\! [\,\hat{z}_{k,0}^{\!\top\!}\,\cdots\,\hat{z}_{k,L-1}^{\!\top\!}\,]^{\top}$ and $\hat{w}_k \!=\! [\,\hat{w}_{k,0}^{\!\top\!}\,\cdots\,\hat{w}_{k,L-1}^{\!\top\!}\,]^{\top}$ 
are PCE coefficients, $\mathbf{\hat{F}}$ has $\sum_{i=0}^{L-1} \mathbf{\hat{F}}_i \Psi_{ijr}$ as its $j,r$-th partition, where $\Psi_{ijr} \!=\! \E(\Phi_i\Phi_j\Phi_r) /\E(\Phi_j^2)$, $j,r \!\in\! \{0$,$1$,$\cdots$,$L-1\}$, $\mathbf{\hat{G}}\!=\![\,\mathbf{\hat{G}}_0^{\!\top\!}~\mathbf{\hat{G}}_1^{\!\top\!} ~\cdots~\mathbf{\hat{G}}_{L-1}^{\!\top\!}\,]^{\!\top\!}$, and $\mathbf{H}\!=\!\mathrm{diag}(\,\mathbf{\hat{H}}_0\,\mathbf{\hat{H}}_1\,\cdots\,\mathbf{\hat{H}}_{L-1}\,)$. The \gls{pce}-based model in Eq.~\eqref{eq:deter_dy} is a generic form of the \gls{pce}-based vehicle model in Appx.-C. 

Compared to the original stochastic model of the \glspl{ov} in Eq.~\eqref{eq:linear_dy_compact_op}, its PCE form as Eq.~\eqref{eq:deter_dy} renders a higher-dimensional deterministic model for the PCE coefficients of the states of the \glspl{ov}. Such a transformation facilitates the solution of the control problem but increases the computational complexity. Note that a deterministic PCE-based model as Eq.~\eqref{eq:deter_dy} only exists for systems with \textit{epistemic} uncertainty, where $\eta$ has a time-invariant joint distribution $p_{\eta}$. Utilizing this model, one can precisely quantify the statistics of $z_k$ and $w_k$ at any time $k \in \mathbb{N}$. For example, the expected value $\mu_{w_k}$ and the covariance $\Sigma_{w_k}$ of $w_k$ read
\begin{equation}\label{eq:moments}
\textstyle \mu_{w_k} = \hat{w}_{k,0},~\Sigma_{w_k} = \sum_{i=1}^{L-1} \hat{w}_{k,i} \hat{w}_{k,i}^{\!\top\!} \E(\Phi_i^2).
\end{equation}
This renders the following PCE-facilitated solution which can be computed using off-the-shelf tools such as \gls{mip}~\cite{kurtz2022mixed}.
\begin{mdframed}[linewidth=1pt, 
backgroundcolor=lightgray!20  
        ]
        \textbf{PCE-Facilitated Control Solution:}
For any time $k \in \mathcal{N}$ with historical measurements $y_0y_1\cdots y_k \!\in\! \mathbf{Y}^+$, solve the following deterministic optimization problem,
\begin{subequations}\label{eq:opti_det}
\begin{alignat}{8}
&\textstyle \min_{\tilde{\useq}_k}  \sum_{i=0}^{N-k-1} \tilde{u}_{i|k}^{\top}R \tilde{u}_{i|k} \\
\mathrm{s.t.}~&\tilde{x}_{i+1|k} \!= A \tilde{x}_{i|k} + B \tilde{u}_{i|k},\\
&\tilde{u}_{i|k} \in \mathbf{U},~\forall \, i\!\in\! \{0,1,\cdots,N\!-\!k\!-\!1\},\\
&\tilde{y}_{i+k|k} \!= C \tilde{x}_{i|k},~\forall \, i\in \{0,1,\cdots,N-k\}, \\
& \tilde{y}_{r|k} = y_r,~\forall \, r\in \{0,1,\cdots,k\}, \label{eq:histequ} \\
& \hat{z}_{i+1} \!= \mathbf{\hat{F}} \hat{z}_i + \mathbf{\hat{G}} \tau_i,~\forall \, i\!\in\! \mathcal{H}, \\
& \hat{w}_i  \!= \mathbf{\hat{H}} \hat{z}_i,~\forall \, i\!\in\! \mathcal{H} \cup \{N\}, \\
& \pmb{\omega}_k \vDash \psi, \label{eq:formalconst}
\end{alignat}
\end{subequations}
where $\pmb{\omega}_k\!:=\!\omega_{0|k} \omega_{1|k} \cdots \omega_{N|k}$ is a deterministic signal with $\omega_{i|k} \!:=\! [\,\tilde{y}_{i|k}^{\top}~\hat{w}_i^{\top}\,]^{\top}$, for $i \!\in\! \mathcal{H} \cup \{N\}$ and $\psi$ is an \gls{stl} formula (see Appx.-A) ensuring $\pmb{\omega}_k \vDash \psi \rightarrow (\tilde{\yseq}_k,\wseq) \vDash \varphi$, which can be straightforwardly determined by substituting the approximation in Eq.~\eqref{eq:moments} to each belief-space predicate of the \gls{dstl} formula $\varphi$.
\end{mdframed}

\begin{remark}[Computational Complexity]\label{rm:compu}
The computational complexity of Eq.~\eqref{eq:opti_det} for time $k \!\in\! \mathcal{H} \!\cup\! \{N\}$ is in general exponential to the number of binary variables brought up by the constraint Eq.~\eqref{eq:formalconst}, according to~\cite{kurtz2022mixed}. Since all history output variables $\tilde{y}_{r|k}$ for $r \!\in\! \{0,1,\cdots,k\}$ are constrained by equality Eq.~\eqref{eq:histequ} without binary variables, the number of binary variables can be estimated by $O(n(N\!-\!k))$, where $n$ is the maximal number of DSTL predicates specified for a specific time from $k$ to $N$, determined by the complexity of the task. Then, the computational complexity of problem Eq.~\eqref{eq:opti_det} at time $k \!\in\! \mathcal{H} \cup \{N\}$ is estimated as $O(2^{n(N-k)})$, leading to a total complexity as $O(\sum_{k=0}^{N}2^{n(N-k)})\!=\!O(\frac{2^{nN}-1}{2^n-1})\!<\!O(2^{n(N-1)})$. Thus, the computational complexity of the intention-aware controller highly depends on the number of DSTL predicates $n$ and the length of the control horizon $N$.
\end{remark}

Computational complexity has been a common and open problem for formal control methods. To solve this problem, approaches like specification decomposition \cite{leahy2022fast} and timing split~\cite{zhang2023modularized} have been proposed to reduce the task complexity (quantified by $n$ in Remark~\ref{rm:compu}) and horizon ($N$ in Remark~\ref{rm:compu}), respectively. Specification decomposition aims at decomposing temporal logic formulas into the ones defined for signals with reduced dimensions. Timing split attempts to separate temporal logic formulas along the timing horizon. Given these methods, resolving computational complexity for intention-aware control can be achieved by decomposing DSTL specifications into smaller and shorter formulas, which is beyond the scope of this work and will be studied in future work.

\section{Experimental Studies}\label{sec:sim}

This section uses two representative autonomous driving cases, namely an overtaking case and an intersection case, to show the comprehensive efficacy of our intention-aware control method. The overtaking case renders a relatively simple setting but allows multiple contexts for distinguished driving intentions. We use this case to demonstrate how the proposed controller ensures automatic adaptation to diverse intentions. Meanwhile, the intersection case originally brought up in Sec.~\ref{sec:intro} renders a more realistic scenario including both vehicles and pedestrians. Although simplified models are used to ensure feasibility and to limit computational complexity, it is sufficiently complicated and comprehensive to verify the applicability of the proposed method in a representative scenario. A comparison study is also provided to showcase the advantage of an intention-aware controller compared to the conventional no-awareness controller.

All experiments are programmed in Python, solved using the \textit{stlpy}~\cite{kurtz2022mixed} and the \textit{chaospy}~\cite{feinberg2018multivariate} libraries, and run on a ThinkPad Laptop with an Intel\textregistered~Core i7-10750H CPU. 
The code can be found at \url{https://zenodo.org/records/11274822}. A video demonstration of the experimental studies is available at \url{https://youtu.be/Pvaj_e2hlY0?si=AlIOeVMFHIrIjnQI}.

\subsection{Case I: Overtaking}\label{sec:cs_1}

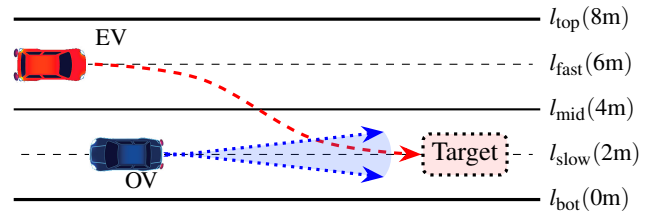
\begin{figure}[htbp]
\noindent
\hspace*{\fill} 
\begin{tikzpicture}[scale=0.8]

\definecolor{sblue}{RGB}{51, 100, 255}
\definecolor{sred}{RGB}{255, 229, 229}

\draw[color=black, very thick] (0, 0) -- (7, 0);
\draw[color=black, dashed] (0.1, 0.6) -- (6.9, 0.6);
\draw[color=black, thick] (0, 1.2) -- (7, 1.2);
\draw[color=black, dashed] (0.1, 1.8) -- (6.9, 1.8);
\draw[color=black, very thick] (0, 2.4) -- (7, 2.4);

\node[anchor=west] () at (7, 0) {\small $l_{\mathrm{bot}}(0\mathrm{m})$};
\node[anchor=west] () at (7, 0.6) {\small $l_{\mathrm{slow}}(2\mathrm{m})$};
\node[anchor=west] () at (7, 1.2) {\small $l_{\mathrm{mid}}(4\mathrm{m})$};
\node[anchor=west] () at (7, 1.8) {\small $l_{\mathrm{fast}}(6\mathrm{m})$};
\node[anchor=west] () at (7, 2.4) {\small $l_{\mathrm{top}}(8\mathrm{m})$};

\node[inner sep=2pt, anchor=center] (ego) at (0.5cm, 1.8cm) {\includegraphics[width=1cm]{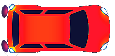}};
\node[inner sep=2pt, anchor=center] (oppo) at (1.5cm, 0.6cm){\includegraphics[width=1cm]{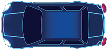}};

\node[anchor=south] (egoc) at ([xshift=0.8cm, yshift=-0.4cm] ego.north) {\small EV};
\node[anchor=north] (oppoc) at ([xshift=-0.2cm, yshift=0.2cm] oppo.north west) {\small OV};

\node[minimum height=0.4cm, minimum width=1cm, draw, fill=sred, dotted, very thick, rounded corners=0.1cm] (target) at (6cm,0.6cm) {Target};
%\node[anchor=west] (ec) at (target.north east) {Target};

\draw [->, >=Stealth, very thick, red, dashed] plot [smooth, tension=0.6] coordinates { (ego.east) (2.5cm, 1.6cm) (4cm, 0.8cm) (target.west)};

\path[fill=sblue,opacity=0.2] (oppo.east) -- 
plot [smooth, tension=0.8] coordinates { (oppo.east) (2.5cm, 0.63cm) ([xshift=-0.5cm, yshift=0.3cm] target.west)} -- 
plot [smooth, tension=1.2] coordinates { ([xshift=-0.5cm, yshift=0.3cm] target.west) ([xshift=-0.4cm] target.west) ([xshift=-0.5cm, yshift=-0.3cm] target.west)} -- 
plot [smooth, tension=0.8] coordinates {  ([xshift=-0.5cm, yshift=-0.3cm] target.west) (2.5cm, 0.57cm) (oppo.east)} -- cycle;

\draw [->, >=Stealth, very thick, blue, dotted] plot [smooth, tension=0.8] coordinates { (oppo.east) (2.5cm, 0.63cm) ([xshift=-0.5cm, yshift=0.3cm] target.west)};

\draw [->, >=Stealth, very thick, blue, dotted] plot [smooth, tension=0.8] coordinates { (oppo.east) (2.5cm, 0.57cm) ([xshift=-0.5cm, yshift=-0.3cm] target.west)};

\end{tikzpicture}
\hspace{\fill} 
\caption{The overtaking case, where an \gls{ev} in the fast lane should switch to the slow lane and overtake an \gls{ov} within a finite time (the dashed arrow). The area in front of the \gls{ov} bounded by two dotted arrows denotes the joint epistemic uncertainty at a certain risk level.}
\label{fig:epist}
\end{figure}

We use an overtaking case to show that an intention-aware controller allows an agent to automatically adjust its behavior according to the intents of its opponents. As shown in Fig.~\ref{fig:epist}, an \gls{ev} needs to overtake an \gls{ov} without collisions. The \gls{ev} and the \gls{ov} are described as deterministic and stochastic linearized bicycle models (see Appx.-B and -C), respectively. The \gls{ov} has a random length $l$ and a random control offset $\delta$ (see Appx.-C), both sampled from non-Gaussian distributions, with $l$ from a uniform distribution $\mathcal{U}(3.99, 4.01)$ and $\delta$ from a truncated Gaussian distribution $\mathcal{N}( 0, 0.1^2 )_{[\,-0.1, 0.1]}$. We consider three different scenarios where the \gls{ov} intends to \textit{slow down}, \textit{speed up}, or \textit{switch to the fast lane}, corresponding to \textit{cooperative}, \textit{antagonistic}, and \textit{ignorant} driving styles, respectively. 

The overtaking task is first described as natural language. \textit{The \gls{ev} eventually switches to the slow lane and drives in front of the \gls{ov} at a maximal risk level $0.05$ (95\% success probability), if the \gls{ov} does not switch to the fast lane or exceed a speed limit $v_{\mathrm{lim}}\!=\!12\,$m/s (in an average sense). Meanwhile, the \gls{ev} should always drive inside the lanes and keep a minimal distance with the \gls{ov}, $d_{\mathrm{safe},1}\!=\!4\,$m longitudinally and $d_{\mathrm{safe},2}\!=\!2\,$m latitudinally, at a maximal risk level $0.05$ (95\% safe).} 
Then, the task is specified using a \gls{dstl} formula for the stochastic model of the \gls{ov} and converted to an \gls{stl} formula for its PCE model (See Appx.-D). We solve the \gls{mpc} problem in Eq.~\eqref{eq:opti_det} with a discrete-time $\Delta t\!=\!1$s and a finite control horizon with length $N\!=\!15$. The control cost is set as $R\!=\!\mathrm{diag}(10^4, 10^{-6})$ to penalize over-steering and encourage speed adjustment. Each time we solve Eq.~\eqref{eq:opti_det}, we update the linearized vehicle models using their current states to reduce the impact of linearization errors. 

\begin{figure}[htbp]
\centering
\subfloat[\footnotesize Assuming the \gls{ov} slowing down, the \gls{ev} performs overtaking.]{\includegraphics[width=0.45\textwidth]{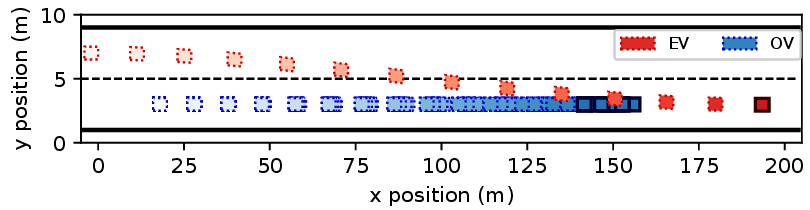}%
\label{fig:exp_image_1}}
\hfil
\subfloat[\footnotesize Assuming the \gls{ov} speeding up, the \gls{ev} does not overtake.]{\includegraphics[width=0.45\textwidth]{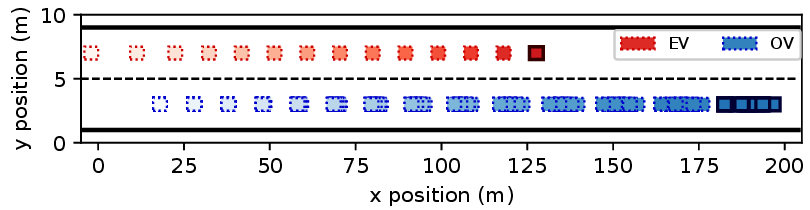}%
\label{fig:exp_image_2}}
\hfil
\subfloat[\footnotesize Assuming the \gls{ov} switching lane, the \gls{ev} slows down.]{\includegraphics[width=0.45\textwidth]{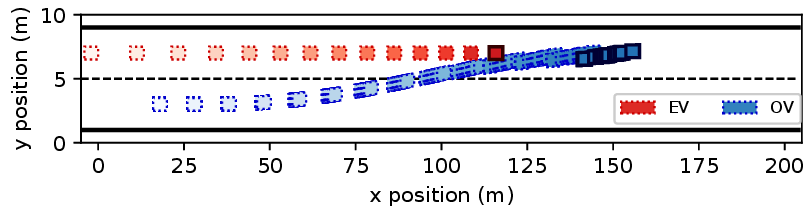}%
\label{fig:exp_image_3}}
\caption{The trajectory of the \gls{ev} and sampled trajectories of the \gls{ov}.}
\label{fig:uc1_result}
\end{figure}

The resulting trajectories of the \gls{ev} and the \gls{ov} with various intentions are shown in Fig.~\ref{fig:uc1_result}. We can see that the \gls{ev} changes its behaviors when the intention of the \gls{ov} varies. As shown in Fig.~\ref{fig:exp_image_1}), when the \gls{ov} intends to keep a constant speed and is not expected to exceed the speed limit $v_{\mathrm{lim}}$, the \gls{ev} performs a successful overtake. When the \gls{ov} intends to speed up as to exceed the speed limit $v_{\mathrm{lim}}$, as shown in Fig.~\ref{fig:exp_image_2}), the \gls{ev} gives up overtaking and maintains its current speed. When the \gls{ov} intends to perform an antagonistic lane-switching, as shown in Fig.~\ref{fig:exp_image_3}), the \gls{ev} slows down and remains in the fast lane to avoid potential collisions. 100 \gls{ov} trajectories sampled from a Monte Carlo study imply zero collision cases, indicating a risk level lower than 0.05. Thus, the behavior of the \gls{ev} satisfies the overtaking task specification for all three scenarios despite the linearization errors, implying the efficacy of the proposed intention-aware control method.

\begin{figure}[htbp]
     \centering
     \includegraphics[width=0.36\textwidth]{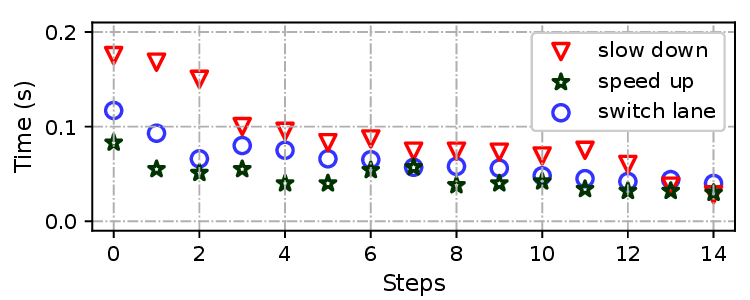}
\caption{The computation time per step (Case I).}
     \label{fig:uc1_complexity}
\end{figure}

The computation time per step for all three scenarios is presented in Fig.~\ref{fig:uc1_complexity}, showing a clear exponentially decaying trend consistent with our analysis in Remark~\ref{rm:compu}. The \textit{slow down} (for \gls{ov}) scenario involves heavier computations than the other two since it is the only one that leads to successful \gls{ev} overtaking. Overall, the overtaking task is sufficiently simple such that all computations have been accomplished within 0.2\,s, much smaller than the discrete sampling time 1\,s. This indicates that our method is promising to be deployed to practical vehicles with real-time performance requirements.

%In each scenario of this case study, we assume that the \gls{ov} has a certain and unchanged intention known to the \gls{ev}, which is not always practical in reality. In practice, the \gls{ev} usually need to infer the \gls{ov}'s intention from its history behavior~\cite{xing2019driver}. Moreover, practical vehicles may reveal different intentions in sequence. In such cases, the \gls{ev} should not only infer the sequential intentions of the \gls{ov} but also estimate the time horizons of these intentions. This renders an interesting but challenging sequential intention-aware control problem which is expected to be investigated in future work.

\subsection{Case II: Intersection}\label{sec:intersec}

In this study, we use the intersection case raised in Sec.~\ref{sec:intro} (See Fig.~\ref{fig:intersection}) to show how an intention-aware controller ensures risk restriction for uncertain intentions. 
Like Sec.~\ref{sec:cs_1}, the \gls{ev} is described using a deterministic linearized bicycle model. The \gls{ov} is a stochastic model with an unknown length $l$ and a random control offset $\delta$ sampled from a uniform distribution $\mathcal{U}(3.59, 3.61)$ and a Gaussian distribution $\mathcal{N}(0, 0.01^2 )$, respectively. Also, we use a similar stochastic vehicle model, with an unknown length sampled from $\mathcal{U}(0.490, 0.501)$ and a random control offset sampled from $\mathcal{N}(0, 0.01^2 )$, to denote the distributional behavior of a group of pedestrians. We consider an uncertain intention of the \gls{ov} with possible intents $\{$\textit{speed~up}, \textit{constant~speed}, \textit{slow~down}$\}$, corresponding to \textit{antagonistic}, \textit{ignorant}, and \textit{cooperative} driving strategies, respectively. This intention is subject to a uniform distribution $\{1/3, 1/3, 1/3\}$. Also, we assume that the pedestrians have an uncertain intention with intents $\{$\textit{move~fast}, \textit{move~slowly}$\}$ subject to a uniform distribution $\{1/2, 1/2\}$. See Appx.-E for the detailed experimental configurations.

The turning task is interpreted as: \textit{The \gls{ev} eventually reaches the target lane. Besides, the \gls{ev} should only turn beyond the first pedestrian crossing and should always avoid collisions with the \gls{ov} and the pedestrians at a risk level $0.05$.} Then, the task is specified using a \gls{dstl} formula and converted to an \gls{stl} formula (See Appx.-E) to facilitate a deterministic \gls{mpc} problem as in Eq.~\eqref{eq:opti_det}. We consider a discrete-time $\Delta t\!=\!0.5\,$s, a finite horizon $N\!=\!36$, and a control cost $R\!=\!\mathrm{diag}(1, 50)$. Similar to Case I, we update the linearized vehicle models using the most recent measurements before solving the optimization problem Eq.~\eqref{eq:opti_det} at each time $k \!\in\! \{0,1,\cdots, N-1\}$ to mitigate the influence of linearization errors.

\begin{figure}[htbp]
\centering
\subfloat[\footnotesize Step $16$.]{\includegraphics[width=0.2\textwidth]{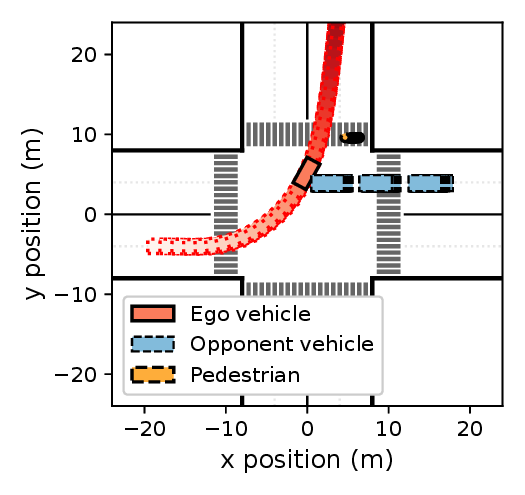}%
\label{fig:uc2_1_1}}
\hfil
\subfloat[\footnotesize Step $20$.]{\includegraphics[width=0.2\textwidth]{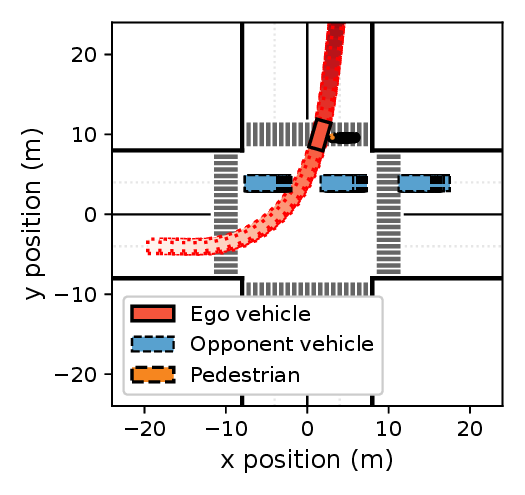}%
\label{fig:uc2_1_2}}
\hfil
\subfloat[\footnotesize Step $16$.]{\includegraphics[width=0.2\textwidth]{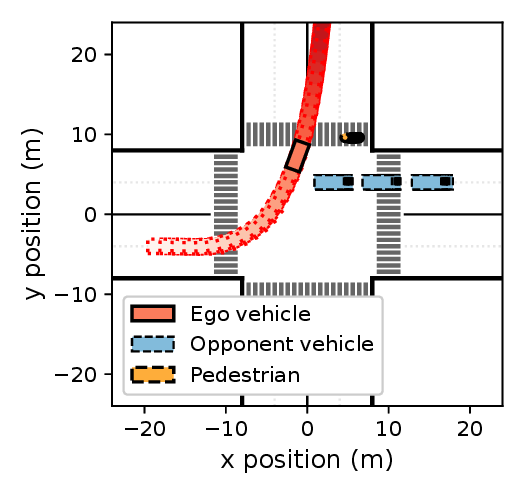}%
\label{fig:uc2_2_1}}
\hfil
\subfloat[\footnotesize Step $20$.]{\includegraphics[width=0.2\textwidth]{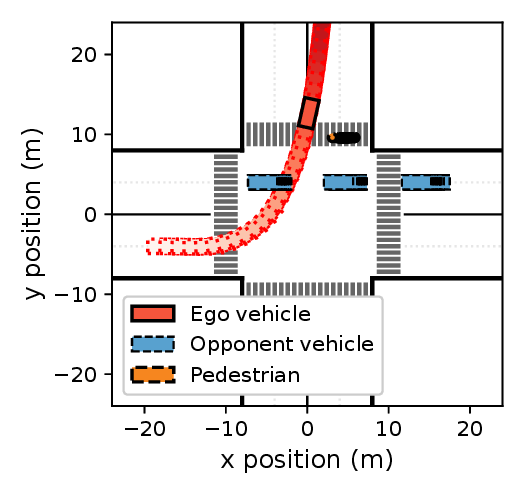}%
\label{fig:uc2_2_2}}
\caption{Results of comparison study. (a) and (b): no awareness; (c) and (d): with intention awareness.}
     \label{fig:uc2}
\end{figure}

We use a comparison study to address the advantage of intention-aware control in risk restriction over a general no-awareness control method. For the former, we consider a specification incorporating the uncertain intentions of the \gls{ov} and the pedestrians. We use the same specification for the latter but omit the uncertain intentions (See Appx.-E for details). 
The resulting trajectories of the vehicles and the pedestrians are shown in Fig.~\ref{fig:uc2}. The trajectories of the \gls{ov} and the pedestrian are the results of 100 times Monte Carlo sampling, implying two multi-modal distributions, with each modal denoting their possible positions given an assumed intent. We can see that the \textit{no-awareness} \gls{ev} is at risk of colliding with the \gls{ov} for the \textit{speeding-up} intent at time step $k\!=\!16$. 
The multi-modal characteristics of the pedestrian are not obvious due to its low speed. But we can still see that \textit{no-awareness} \gls{ev} is at risk of colliding with the \textit{moving-fast} pedestrians at time step $k\!=\!20$. However, the intention-aware \gls{ev} has successfully avoided collisions with all possible positions of the \gls{ov} and the pedestrians by turning slightly earlier compared to the no-awareness case (See Fig.~\ref{fig:uc2}). Although this strategy seems antagonistic and aggressive from a practical perspective, it restricts the risk under the predefined level of $0.05$. 

The results indicate that the proposed intention-aware controller ensures superior safety compared to a conventional no-awareness controller. In fact, the latter only considers the average behaviors of the opponents while the former incorporates all their possible behaviors. This means that intention-aware control can better understand the worst-case behaviors of the opponents, leading to an advantageous capability of reasoning about risk. Using PCE allows the precise characterization of nontrivial uncertainties raised by incorporating intentions. As a result, intention-aware control can guarantee provably restricted risk, even when the opponents perform antagonistic or adversarial behaviors. However, intention-aware control can only reason about the risk of the intentions incorporated in a predefined set. It is not effective for unexpected behaviors of the opponents. In unexpected cases, the ego vehicle should take the worst-case response, such as an emergency stop or driving to a safe zone, to ensure safety.

\begin{figure}[htbp]
     \centering
     \includegraphics[width=0.36\textwidth]{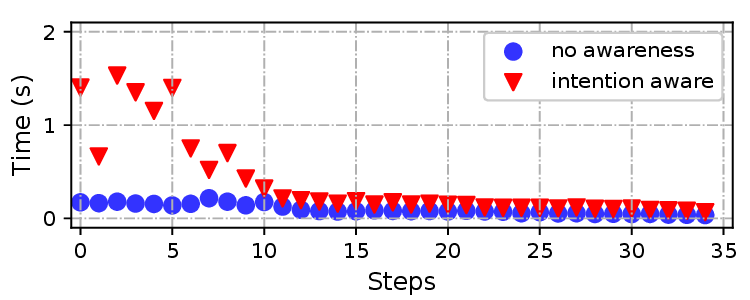}
\caption{The computation time per step (Case II).}
     \label{fig:uc2_complexity}
\end{figure}

The computation times per step for both \textit{no-awareness} and \textit{intention-awareness} scenarios are presented in Fig.~\ref{fig:uc2_complexity}. Both scenarios show exponentially decaying computation times as the step increases, implying the correctness of our analysis on computational complexity in Remark~\ref{rm:compu}. It is noticed that the intention-aware controller involves heavier computation than the \textit{no-awareness} one due to the incorporation of uncertain intentions. The computational load is substantial before step $10$ when the \gls{ev} tries to find a feasible solution to avoid collisions with the \gls{ov} and the pedestrians. The computation time before step $8$ even exceeds the discrete sampling time $0.5\,$s, bringing up a challenge to its deployment on a real-time system. Possible solutions to resolve this problem may include decomposition and splitting methods as interpreted in Sec.~\ref{sec:solution}.

\section{DISCUSSION}\label{sec:disc}

The most important result of this paper is a tractable and efficient PCE-facilitated solution for the intention-aware control of an autonomous vehicle with uncertain opponent traffic participants. To derive this, we describe the opponent participants as stochastic dynamic models with epistemically uncertain intentions and parameters. This allows us to characterize the random behaviors of these opponents into a deterministic PCE-based model. Through a novel DSTL belief-space specification, we convert a challenging intention-aware control problem into a deterministic MPC problem that can be easily solved using off-the-shelf tools. Besides, the controller is correct by design, with ensured provable risk restriction. Theoretical analysis and experimental studies have implied its promising real-time implementation. Finally, the experimental results have sketched an interactive control strategy that guarantees superior safety and flexibility in complex driving environments compared to the conventional no-awareness controllers.

Our work has two main limitations that are worth studying in future work. Firstly, our method assumes that the stochastic characteristics of the uncertainty and intentions of all opponents are known. In practice, however, the stochastic characteristics of the intention and parameters of the opponents may change and may even be unknown to the ego vehicle. In this case, the characteristics of the uncertainties and intentions should be learned and updated using the historical observations of the opponents. This can significantly enhance the prediction accuracy of the opponents' intentions, thus improving the robustness and adaptability of the intention-aware controller. To achieve this, the driving tasks can be specified on a belief space that incorporates the historical observations of the opponents. Then, Bayesian inference~\cite{shao2023does} and Kalman filter~\cite{schulz2018multiple} can be used to update the characteristics of the uncertainties and learn the intentions using historical observations. Secondly, the current solution still suffers from the heavy computational complexity for complicated scenarios. Specification decomposition and timing splitting methods could be considered to reduce the computational complexity, as elucidated in Sec.~\ref{sec:solution}.

\section{CONCLUSIONS}\label{sec:con}

This paper provides the first correct-by-design intention-, interaction-, and risk-aware controller for autonomous vehicles by characterizing complex epistemic uncertainty using stochastic expansion. It renders an efficient solution that can be solved using off-the-shelf tools. In the future, we will study the extension of this work to risk-restriction control incorporating observations of the opponents. We will also investigate the reduction of computational complexity of the method.

%\addtolength{\textheight}{-12cm}   % This command serves to balance the column lengths
                                  % on the last page of the document manually. It shortens
                                  % the textheight of the last page by a suitable amount.
                                  % This command does not take effect until the next page
                                  % so it should come on the page before the last. Make
                                  % sure that you do not shorten the textheight too much.

%%%%%%%%%%%%%%%%%%%%%%%%%%%%%%%%%%%%%%%%%%%%%%%%%%%%%%%%%%%%%%%%%%%%%%%%%%%%%%%%

%%%%%%%%%%%%%%%%%%%%%%%%%%%%%%%%%%%%%%%%%%%%%%%%%%%%%%%%%%%%%%%%%%%%%%%%%%%%%%%%

%%%%%%%%%%%%%%%%%%%%%%%%%%%%%%%%%%%%%%%%%%%%%%%%%%%%%%%%%%%%%%%%%%%%%%%%%%%%%%%%%

\section*{APPENDIX}

\subsection{Signal Temporal Logic (STL)}

For a discrete-time signal $\sseq:=s_0 s_1\cdots$, where $s_k \in \mathbb{R}^n$ for $k \in \mathbb{N}$ and $n \in \mathbb{Z}^+$, the syntax of \gls{stl} is given as~\cite{salamati2021data}
\begin{equation*}
\psi ::= \top\,|\, \mu \,|\, \lnot \psi \,|\, \psi_1 \!\wedge\! \psi_2 \,|\, \psi_1 \mathsf{U}_{[\,a,\,b\,]} \psi_2,
\end{equation*}
where $\psi_1$, $\psi_1$, and $\psi_2$ are \gls{stl} formulas, 
$\mu$ is a predicate with a mapping $h: \mathbb{R}^n \!\rightarrow\! \mathbb{R}$ 
via $\displaystyle \mu\!:=\! \left\{ \begin{array}{ll}
\!\!\!\top, &h(s_k) \geq 0 \\
\!\!\!\bot, &h(s_k) < 0
\end{array} \right.$, $k\!\in\!\mathbb{N}$. The \textit{disjunction}, \textit{eventually}, and \textit{always} operators are defined as $\psi_1 \!\vee\! \psi_2 \!:=\! \lnot \left(\lnot \psi_1 \wedge \lnot \psi_2 \right)$, 
$\mathsf{F}_{[\,a,\,b\,]} \psi \!:=\! \top \mathsf{U}_{[\,a,\,b\,]} \psi$,
$\mathsf{G}_{[\,a,\,b\,]} \psi\!:=\! \lnot \left( \top \mathsf{U}_{[\,a,\,b\,]} \lnot \psi \right)$. The semantics of STL denoted as $(\sseq,k) \vDash \psi$ are recursively determined as 
\begin{equation*}
\begin{split}
&(\sseq,k) \vDash \mu \! \leftrightarrow \! h(s_k)\!\geq\!0,~ 
(\sseq,k) \vDash \lnot \psi \! \leftrightarrow \! \lnot ((\sseq,k) \vDash \psi), \\
&(\sseq,k) \vDash \psi_1 \wedge \psi_2 \! \leftrightarrow \! (\sseq,k) \vDash \psi_1 \wedge (\sseq,k) \vDash \psi_2, \\
&(\sseq,k) \vDash \psi_1 \vee \psi_2 \! \leftrightarrow \! (\sseq,k) \vDash \psi_1 \vee (\sseq,k) \vDash \psi_2, \\
&(\sseq,k) \vDash \mathsf{F}_{[\,a,\,b\,]} \psi \! \leftrightarrow \! \exists \, k' \!\in\! [\,k\!+\!a,\,k\!+\!b\,],~\mathrm{s.t.}, (\sseq, k') \vDash \psi, \\
&(\sseq,k) \vDash \mathsf{G}_{[\,a,\,b\,]} \psi \! \leftrightarrow \! \forall \, k' \!\in\! [\,k\!+\!a,\,k\!+\!b\,],~\mathrm{s.t.}, (\sseq, k') \vDash \psi, \\
&(\sseq,k) \vDash \psi_1 \mathsf{U}_{[\,a,\,b\,]} \psi_2 \! \leftrightarrow \! \exists \, k' \!\in\! [k\!+\!a,k\!+\!b],~\mathrm{s.t.}, (\sseq, k') \vDash \psi_2, \\
&~~~~~~~~~~~~~~~~~~~~~~~~~~\mathrm{and}~\forall \, k'' \in [\,k, \,k'\,], (\sseq,k'') \vDash \psi_1.
\end{split}
\end{equation*}
For $k=0$, the symbol $k$ can be omitted, rendering  $\sseq \vDash \psi$.
The robustness of an \gls{stl} formula $\psi$ denoted as $\rho^{\psi}(\sseq,k)$, with $\rho^{\psi}(\sseq,k)>0 \leftrightarrow (\sseq, k) \vDash \varphi$, is inductively defined as
\begin{equation*}
\begin{split}
&\textstyle \rho^{\mu}(\sseq,k) \!:=\! h(s_k),~
\rho^{\lnot \psi}(\sseq,k) \!:=\! - \rho^{\psi}(\sseq,k),\\
&\textstyle \rho^{\psi_1 \wedge \psi_2}(\sseq,k)\!: =\! \min(\rho^{\psi_1}(\sseq, k),\, \rho^{\psi_2}(\sseq,k)), \\
&\textstyle \rho^{\psi_1 \vee \psi_2}(\sseq,k) \!: =\! \max(\rho^{\psi_1}(\sseq,k),\, \rho^{\psi_2}(\sseq,k)), \\
&\textstyle \rho^{\!\psi_1 \!\mathsf{U}_{[\,a,\,b\,]}\! \psi_2\!}(\sseq,k) := \max_{k' \in [\,k+a,\,k+b\,]} 
(\min ( \rho^{\psi_2}(\sseq, k'),\\
&\textstyle ~~~~~~~~~~~~~~~~~~~~~~~~~~\min_{k'' \in [\,k, \,k'\,]} \rho^{\psi_1}(\sseq, k'') ) ),  \\ 
&\textstyle \rho^{\mathsf{F}_{[\,a,\,b\,]} \psi}(\sseq, k ) \!:=\! \max_{k' \in [\,k+a,\,k+b\,]} \rho^{\psi}(\sseq, k'), \\
&\textstyle \rho^{\mathsf{G}_{[\,a,\,b\,]} \psi}(\sseq,k)\!:=\min_{k' \in [\,k+a,\,k+b\,]} \rho^{\psi}(\sseq, k'). 
\end{split}
\end{equation*}
For $k=0$, the robustness reads $\rho^{\psi}(\sseq)$.

\subsection{Describing EV Using a Deterministic Linear Model}

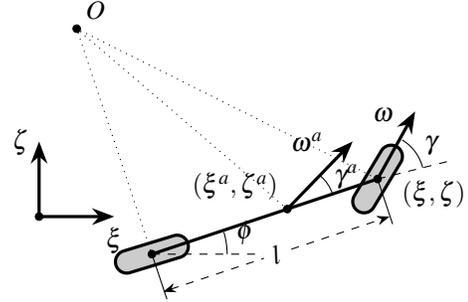
\begin{figure}[htbp]
\noindent
\begin{center}
\begin{tikzpicture}[scale=1]

\definecolor{sgray}{RGB}{204, 204, 204}

\node[circle, inner sep=0pt, minimum size=1mm, draw, fill=black] (origin) at (-0.5cm, 0.5cm) {};
\draw[->, >=Stealth,color=black, very thick] (origin.center) --node[pos=1, anchor=north]{$\xi$} ([xshift=1cm] origin.center);
\draw[->, >=Stealth,color=black, very thick] (origin.center) -- node[pos=1, anchor=east]{$\zeta$} ([yshift=1cm] origin.center);

\node[circle, inner sep=0pt, minimum size=1mm, draw, fill=black] (O) at (0cm, 3cm) {};
\node[anchor=south west] (On) at (O) {$O$};

\node[minimum height=0.3cm, minimum width=1cm, draw, very thick, fill=sgray, rounded corners=0.15cm, rotate=17, anchor=center] (back_tire) at (1cm,0cm) {};
\node[minimum height=0.3cm, minimum width=1cm, draw, very thick, fill=sgray, rounded corners=0.15cm, rotate=59, anchor=center] (front_tire) at (4cm,1cm) {};

\draw[color=black, very thick] (back_tire.center) -- (front_tire.center);
\node[circle, inner sep=0pt, minimum size=1mm, draw, fill=black] (back_tire_c) at (back_tire.center) {};
\node[circle, inner sep=0pt, minimum size=1mm, draw, fill=black] (front_tire_c) at (front_tire.center) {};
\node[circle, inner sep=0pt, minimum size=1mm, draw, fill=black] (C) at (2.8cm, 0.6cm) {};
\draw[color=black, dotted] (O) -- (back_tire.center);
\draw[color=black, dotted] (O) -- (front_tire.center);
\draw[color=black, dotted] (O) -- (C.center);
\draw[->, >=Stealth,color=black, very thick] (C.center) -- node[pos=0.7, anchor=south east]{$\omega^a$} ([xshift=0.87cm, yshift=0.9cm] C.center);
\draw ([xshift=0.6cm, yshift=0.2cm] C.center) arc[start angle=17, end angle=50, x radius=0.5cm, y radius=0.5cm];
\node[anchor=south west] (gamma) at (3.3cm, 0.75cm) {$\gamma^a$};

\draw[color=black, dashed] (front_tire.center) -- ([xshift=1cm, yshift=0.25cm] front_tire.center);
\draw[->, >=Stealth,color=black, very thick] (front_tire.center) -- node[pos=0.7, anchor=south east]{$\omega$}([xshift=0.5cm, yshift=0.9cm] front_tire.center);
\draw ([xshift=0.6cm, yshift=0.15cm] front_tire.center) arc[start angle=14, end angle=59, x radius=0.6cm, y radius=0.6cm];
\node[anchor=south west] (gamma) at (4.5cm, 1.2cm) {$\gamma$};

\draw[color=black, dashed] (back_tire.center) -- ([xshift=1.5cm] back_tire.center);
\draw ([xshift=1cm, yshift=0cm] back_tire.center) arc[start angle=0, end angle=17, x radius=1cm, y radius=1cm];
\node[anchor=south west] (gamma) at (2cm, 0cm) {$\phi$};

\node[anchor=north west] (gamma) at ([xshift=0.2cm, yshift=0.1cm] front_tire.center) {$(\xi, \zeta)$};
\node[anchor=south east] (gamma) at (C.center) {$(\xi^a, \zeta^a)$};

\draw[color=black] (back_tire.center) -- ([xshift=0.2cm, yshift=-0.6cm] back_tire.center);
\draw[color=black] (front_tire.center) -- ([xshift=0.2cm, yshift=-0.6cm] front_tire.center);
\draw[<->, >=Stealth,color=black, dashed] ([xshift=0.15cm, yshift=-0.5cm] back_tire.center) -- node[pos=0.5, anchor=center, fill=white, rotate=17]{$l$}([xshift=0.16cm, yshift=-0.5cm] front_tire.center);

\end{tikzpicture}
%\hspace{\fill} 
\end{center}
\caption{The bicycle model for autonomous vehicles.}
\label{fig:bicycle}
\end{figure}

An autonomous vehicle is typically described as a nonlinear bicycle model~\cite{kong2015kinematic}. We use a deterministic bicycle model illustrated in Fig.~\ref{fig:bicycle} to describe the behavior of the \gls{ev}, 
\begin{equation}\label{eq:bicycle}
\mathrm{\gls{ev}}: \left\{ \begin{array}{l}
\xi_{k+1} \!=\! \xi_k \!+\! \Delta t  \omega_k \cos(\phi_k\!+\!\gamma_k), \\
\zeta_{k+1} \!=\! \zeta_k \!+\! \Delta t  \omega_k \sin(\phi_k\!+\!\gamma_k), \\
\phi_{k+1} \!=\! \phi_k \!+\! \Delta t \omega_k \sin \gamma_k /l,\\
\omega_{k+1} \!=\! \omega_k \!+\! \Delta t a_k, 
\end{array} \right. 
\end{equation}
where $(\xi_k, \zeta_k)$ denote the planar coordinate of the mass center of the front wheel at discrete time $k\in\mathbb{N}$, $\phi_k$ and $\omega_k$ are the heading angle of the vehicle and the linear velocity of the front wheel, respectively, $\gamma_k$, $a_k$ are the steering angle and the linear acceleration of the front wheel, respectively, $\Delta t$ is the discrete sampling time, and $l$ is the length of the vehicle. Different from~\cite{kong2015kinematic}, all variables in Eq.~\eqref{eq:bicycle} are defined for the front wheel instead of the mass center of the vehicle to obtain a simplified model. In fact, it is straightforward to convert~Eq.~\eqref{eq:bicycle} to a model for an arbitrary point on the vehicle. Let $(\xi^a_k, \zeta^a_k)$,  $\phi^a_k$, and $\omega^a_k$ be the planar coordinate, the heading angle, and the linear velocity of any mass point on the vehicle as shown in Fig.~\ref{fig:bicycle}, their relations with the front wheel variables $(\xi_k, \zeta_k)$,  $\phi_k$, and $\omega_k$ are described by $\xi^a_k = \xi_k \!-\! (1\!-\!c)l \cos \phi_k$, $\zeta^a_k = \zeta_k \!-\! (1\!-\!c)l \sin \theta_k$, $\omega^a_k = {\omega_k \cos \gamma_k}/{\cos \gamma^a_k}$, and $\gamma^a_k=\tan^{-1}\! \left(c \tan \gamma_k \right)$, where $c\in [\,0,\,1\,]$ is a ratio scalar such that $cl$ is the distance from this point to the back wheel. 

Let $x_k \!=\! \left[\,\xi_k~\zeta_k~\phi_k~\omega_k\, \right]^{\top}$ be the measurable state of \gls{ev} and $u_k \!=\! \left[\,\gamma_k~a_k\, \right]^{\top}$ be its control input. The deterministic model of the \gls{ev} is compactly represented as
\begin{equation}\label{eq:det_veh}
\mathrm{\gls{ev}}: \left\{ \begin{array}{l}
x_{k+1} \!=\! x_k \!+\! f(x_k, u_k), \\
~~~y_k \!=\! x_k,
\end{array} \right. 
\end{equation}
where $y_k \in \mathbb{R}^4$ is the output of \gls{ev} and
\begin{equation*}
f(x_k, u_k) \!=\! \Delta t \left[\,\omega_k\! \cos(\!\phi_k\!+\!\gamma_k\!)~\omega_k\! \sin(\!\phi_k\!+\!\gamma_k\!)~\frac{\omega_k}{l}\! \sin \gamma_k~a_k\right]^{\top}.
\end{equation*}
By linearizing Eq.~\eqref{eq:det_veh} around $x_0\!=\! [\,\xi_0~\zeta_0~\phi_0~\omega_0\,]^{\!\top}$ and $u_0 \!=\! [\,\gamma_0~a_0\,]^{\!\top}$, we obtain a linearized dynamic model,
\begin{equation*}
f(x_k, u_k) \!=\! f(x_0, u_0) \!+\! A(x_0, u_0)(x_k - x_0) \!+\! B(x_0, u_0)(u_k - u_0),
\end{equation*}
where $A(x_0, u_0) \in \mathbb{R}^{4\times 4}$ and $B(x_0, u_0) \mathbb{R}^{4 \times 2}$ are
\begin{equation*}
\begin{split}
A(x_0, u_0) 
&\!=\! \left[ \begin{array}{cccc}
0 & 0 & -\Delta t  \omega_0 \sin \!\left(\phi_0\!+\!\gamma_0 \right) & \Delta t  \cos\!\left(\phi_0\!+\!\gamma_0 \right) \\
0 & 0 & \Delta t  \omega_0 \cos \!\left(\phi_0\!+\!\gamma_0 \right) & \Delta t   \sin \!\left(\phi_0\!+\!\gamma_0 \right) \\
0 & 0 & 0 & \Delta t  \sin \gamma_0/l \\
0 & 0 & 0 & 0
\end{array} \right],\\
B(x_0, u_0) & \!=\! \left[ \begin{array}{cc}
- \Delta t  \omega_0 \sin \!\left(\phi_0\!+\!\gamma_0 \right) & 0 \\
\Delta t  \omega_0 \cos \!\left(\phi_0\!+\!\gamma_0 \right) & 0 \\
\Delta t  \omega_0 \cos \gamma_0 /l & 0 \\
0 & \Delta t 
\end{array} \right].
\end{split}
\end{equation*}
Assume $\phi_k \approx \phi_0$ for all $k \in \mathbb{N}$ since a vehicle hardly performs large steering. 
With $\gamma_0 = 0$, we rewrite~\eqref{eq:det_veh} as
\begin{equation}\label{eq:li}
\mathrm{\gls{ev}}: \left\{ \begin{array}{l}
x_{k+1} \!=\! A_0x_k \!+\! B_0 u_k \!+\! d_0, \\
~~~y_k \!=\! C_0 x_k,
\end{array} \right.
\end{equation}
which is consistent with the \gls{ev} model in Eq.~\eqref{eq:ego}, where $y_k \in \mathbb{R}^4$ is the output of \gls{ev}, $C_0=I$, and
\begin{equation*}
\begin{split}
A_0 &\!=\! I \!+\! \Delta t\! \left[ \begin{array}{cccc}
0 & 0 & - \omega_0 \sin \phi_0 & \cos \phi_0 \\
0 & 0 & \omega_0 \cos \phi_0 &  \sin \phi_0 \\
0 & 0 & 0 & 0 \\
0 & 0 & 0 & 0
\end{array} \right],\\
B_0 &\!=\! \Delta t\! \left[\!\! \begin{array}{cc}
- \omega_0 \sin \phi_0 & 0 \\
 \omega_0 \cos \phi_0 & 0 \\
0 & 0 \\
0 & 1
\end{array} \!\!\right],
d_0 \!=\! \Delta t\! \left[\!\! \begin{array}{c}
\omega_0 \phi_0 \sin \phi_0 \\
- \omega_0 \phi_0 \cos\phi_0  \\
0 \\
0
\end{array} \!\!\right].
\end{split}
\end{equation*}

\subsection{Describing OV Using a Stochastic Linear Model}

Based on Eq.~\eqref{eq:bicycle}, we use a stochastic bicycle model to describe the behavior of an \gls{ov} with uncertain parameters,
\begin{equation}
\mathrm{\gls{ov}}: \left\{ \begin{array}{l}
\xi_{k+1} \!=\! \xi_k \!+\! \Delta t  \omega_k \cos(\phi_k\!+\!\gamma_k), \\
\zeta_{k+1} \!=\! \zeta_k \!+\! \Delta t  \omega_k \sin(\phi_k\!+\!\gamma_k), \\
\phi_{k+1} \!=\! \phi_k \!+\! \Delta t \omega_k \sin \gamma_k /l,\\
\omega_{k+1} \!=\! \omega_k \!+\! \Delta t (a_k \!+\! \delta), 
\end{array} \right. 
\end{equation}
where $l\sim p_l$ is the unknown length of the vehicle and $\delta \sim p_{\delta}$ is a random acceleration control offset, where $p_l$ and $p_{\delta}$ are predefined \glspl{pdf}.
We assume an integer-valued intention variable $\iota \!\in\! I \!\subset\! \mathbb{Z}$ with a distribution $p_{\iota}$ on $I$ and an intention-encoding feedforward controller $v_k \!=\! \iota \tau_k$ for the \gls{ov}, where $v_k \!=\![\,\gamma_k,\,a_k\,]^{\top}$ is the control input of the \gls{ov} and $\tau_k \!\in\! \mathbb{R}^2$ for $k \!\in\! \mathbb{N}$ is a predefined control signal. 
Supposing that the feedforward control input $\tau_0, \tau_1, \cdots$ is designed such that $v_k = \tau_k$ leads to an accelerating \gls{ov}, the intents $\iota=-1,\,0,\,1$ encode the \textit{slowing-down}, \textit{constant-speed}, and \textit{speeding-up} driving policies, respectively.

Let $z_k \!=\! \left[\,\xi_k~\zeta_k~\phi_k~\omega_k\, \right]^{\top}$ be the measurable state of the \gls{ov}. Similar to Appx.-B, we obtain the following linearized dynamic model around $z_0\!=\! [\,\xi_0~\zeta_0~\phi_0~\omega_0\,]^{\!\top}$,
\begin{equation}\label{eq:li_sta}
\mathrm{\gls{ov}}: \left\{ \begin{array}{l}
\textstyle z_{k+1} \!=\! A_0z_k \!+\!  \sum_{i=0}^1 b_i(\eta) B_i \tau_k \!+\! \sum_{i=0}^1 e_i(\eta) d_i, \\
~~~w_k \!=\!C_0 z_k,
\end{array} \right.
\end{equation}
where $\eta\!=\![\,l~\delta~\iota\,]^{\top}$ is a joint random variable, $b_0(\eta)\!=\!\iota$, $b_1(\eta)\!=\!\iota/{l}$, $e_0(\eta) \!=\! 1$, and $e_1(\eta)\!=\!\delta$, $A_0$, $B_0$, $C_0$, and $d_0$ are defined in Eq.~\eqref{eq:li}, and
\begin{equation}
B_1 = \Delta t \left[ \begin{array}{cc}
0 & 0 \\
0 & 0 \\
\omega_0 \cos \phi_0 & 0 \\
0 & 0
\end{array} \right]~\mathrm{and}
~
d_1 = \Delta t \left[ \begin{array}{c}
0 \\
0 \\
0 \\
1
\end{array} \right].
\end{equation}
The stochastic model Eq.~\eqref{eq:li_sta} has a consistent form with the \gls{ov} model in Eq.~\eqref{eq:linear_dy_compact}.

We use PCE to expand the state and uncertain parameters of Eq.~\eqref{eq:li_sta} as $z_k \!=\! \sum_{j=0}^{L-1} \hat{z}_{k,j} \Phi_j $ for $k \!\in\!\mathbb{N}$, $b_0(\eta) \!=\! \mathbf{\hat{b}}_0{\!\top\!} \pmb{\Phi}$, $b_1(\eta) \!=\! \mathbf{\hat{b}}_1^{\!\top\!} \pmb{\Phi}$, $e_0(\eta) \!=\! \mathbf{\hat{e}}_0^{\!\top\!} \pmb{\Phi}$, $e_1(\eta) \!=\! \mathbf{\hat{e}}_1^{\!\top\!} \pmb{\Phi}$, where $\pmb{\Phi} = [\,\Phi_0,~ \Phi_1,~ \cdots,~ \Phi_{L-1}\,]^{\!\top\!}$ is the polynomial bases of the joint uncertainty $\eta$ with $\Phi_0:=1$, $\hat{z}_{k,j} \in \mathbb{R}^4$ for $j=0,1,\cdots,L-1$ and $\mathbf{\hat{b}}_0, \mathbf{\hat{b}}_1, \mathbf{\hat{e}}_0, \mathbf{\hat{e}}_1\!\in\!\mathbb{R}^L$ are vectors of PCE coefficients. Substituting the expansions above to Eq.~\eqref{eq:li_sta} renders
\begin{equation*}
\begin{split}
\textstyle \sum_{j=0}^{L-1} \hat{z}_{k+1,j} \Phi_j  &=\textstyle  A_0 \sum_{j=0}^{L-1} \hat{z}_{k,j} \Phi_j  \\
+&\textstyle  \sum_{i=0}^1\mathbf{\hat{b}}_i^{\!\top\!} \pmb{\Phi} B_i \tau_k 
+ \sum_{i=0}^1 \mathbf{\hat{e}}_i^{\!\top\!} \pmb{\Phi} d_i.
\end{split}
\end{equation*}
Taking the inner product of both sides with any polynomial bases $\Phi_s$, $s\!\in\!\{0,1,\cdots,L-1\}$, we have
\begin{equation*}
\begin{split}
\textstyle \sum_{j=0}^{L-1} \hat{z}_{k+1,j}&\left\langle\Phi_j , \Phi_s \right\rangle\textstyle \!=\! \sum_{j=0}^{L-1} A_0 \hat{z}_{k,j} \!\left\langle \Phi_j , \Phi_s \right\rangle
 \\
+&\textstyle \sum_{i=0}^1 \mathbf{\hat{b}}_i^{\!\top\!} \!\left\langle\pmb{\Phi}, \Phi_s \right\rangle \!B_i \tau_k \!+\! \sum_{i=0}^1 \mathbf{\hat{e}}_i^{\!\top\!} \!\left\langle\pmb{\Phi}, \Phi_s \right\rangle \! d_i.
\end{split}
\end{equation*}
The orthogonal properties of the bases ensure $\left\langle \Phi_s, \Phi_s \right\rangle \!>\! 0$, $\forall\,s\!\in\!\{0,1,\cdots,N-1\}$ and $\left\langle \Phi_j , \Phi_s \right\rangle \!=\! 0$, $\forall\,j,s\!\in\!\{0,1,\cdots,L-1\}$, $j \!\neq\! s$. 
The PCE model of the \gls{ov} reads
\begin{equation}\label{eq:compactsys}
\begin{split}
\textstyle \mathbf{\hat{z}}_{k+1} &= \mathbf{A} \mathbf{\hat{z}}_k + \mathbf{B} \tau_k + \mathbf{d},\\
 \mathbf{\hat{w}}_k &=  \mathbf{\hat{z}}_k,
\end{split}
\end{equation}
where $\mathbf{\hat{z}}_k=[\,\hat{z}_{k,0}^{\top}~\hat{z}_{k,1}^{\top}~\cdots~\hat{z}_{k,L-1}^{\top}\,]^{\top}$, $\mathbf{A}\!=\!\mathrm{diag}(\underbrace{A_0, \cdots, A_0}_L)$,
\begin{equation*}
\mathbf{B} = \left[ \begin{array}{c}
\textstyle \sum_{i=0}^1 \hat{b}_{i,0} B_i \\
\textstyle \sum_{i=0}^1 \hat{b}_{i,1} B_i \\
\vdots \\
\textstyle \sum_{i=0}^1 \hat{b}_{i,L-1} B_i
\end{array} \right],~\mathrm{and}~\mathbf{d} = \left[ \begin{array}{c}
\textstyle \sum_{i=0}^1 \hat{e}_{i,0} d_i \\
\textstyle \sum_{i=0}^1 \hat{e}_{i,1} d_i \\
\vdots \\
\textstyle \sum_{i=0}^1 \hat{e}_{i,L-1} d_i
\end{array} \right].
\end{equation*}
with an initial condition $\pmb{\hat{z}}_0 = [\,z_0^{\top}~\underbrace{\mathbf{0}_{4\times 1}^{\top}~\cdots~\mathbf{0}_{4\times 1}^{\top}}_{L-1} ]^{\top}$, where $\hat{b}_{i,j}$ and $\hat{e}_{i,j}$ are the $j$-th element of vectors $\mathbf{\hat{b}}_i$ and $\mathbf{\hat{e}}_i$, for $i\!\in\!\{0,1\}$ and $j \!\in\!\{0,1,\cdots,L-1\}$. Eq.~\eqref{eq:compactsys} is consistent with the \gls{pce}-based model of the \glspl{ov} in Eq.~\eqref{eq:deter_dy}.

\subsection{Experimental Configuration of Case I}

Case I considers a deterministic \gls{ev} modeled by Eq.~\eqref{eq:ego} and a stochastic \gls{ov} modeled by Eq.~\eqref{eq:linear_dy_compact_op}. The details of the models are given Appx.-B and Appx.-C. We use a deterministic variable $y_k\!\in\!\mathbb{R}^4$ and a stochastic variable $w_k\!\in\!\mathbb{R}^4$ to represent the outputs of the \gls{ev} and \gls{ov}, respectively. The elements of these variables, i,e., $y_{i,k}$ and $w_{i,k}$ for $i\!\in\!\{1,2,3,4\}$, denote the longitudinal positions, the latitudinal positions, the orientations, and the linear velocities of the vehicles.

Following the natural language description in Sec.~\ref{sec:cs_1}, the overtaking task is specified as \gls{dstl} predicates: \textit{The \gls{ev} eventually switches to the slow lane ($(\nu_1^{\mathrm{E}}, *)\!:=\!|y_{2,k}\!-\!l_{\mathrm{slow}}|\! <\!0.1$m$\,\wedge\,|y_{3,k}| \!<\!0.1$rad) and drives in front of the \gls{ov} at a maximal risk level $0.05$  ($(\nu_2^{\mathrm{E}}, 0.05)\!:=\!P(w_{1,k}\!\leq\! y_{1,k}\!-\! d_{\mathrm{safe},1}) \!\geq\! 0.95 $), if the \gls{ov} does not switch to the fast lane ($(\nu_1^{\mathrm{O}}, 1)\!:=\!\mu_{w_{2,k}}\!<\!l_{\mathrm{mid}}$) or exceed a speed limit $v_{\mathrm{lim}}$ ($(\nu_2^{\mathrm{O}}, 1)\!:=\!\mu_{w_{4,k}}\!<\!v_{\mathrm{lim}}$). Meanwhile, the \gls{ev} should always drive inside the lanes ($(\nu_1^{\mathrm{S}}, *)\!:=\!l_{\mathrm{bot}}\!<\!y_{2,k} \!\leq\! l_{\mathrm{top}}$) and keep a minimal distance with the \gls{ov}, $d_{\mathrm{safe},1}\!=\!10\,$m longitudinally and $d_{\mathrm{safe},2}\!=\!2\,$m latitudinally, at a maximal risk level $0.05$ ($(\nu_2^{\mathrm{S}}, 0.05)\!:=\!P(|w_{1,k} \!-\! y_{1,k}| \!<\! d_{\mathrm{safe},1}\wedge|w_{2,k} \!-\! y_{2,k}| \!<\! d_{\mathrm{safe}, 2}) \!\geq\! 0.95$).} Here, $*$ denotes an arbitrary value in $[\,0,\,1\,]$. Then, this task is specified as $\varphi:=\mathsf{G}_{[\,0,\,N\,]}\bigwedge_{j=1}^2(\nu_j^{\mathrm{S}}, 0.05) \wedge (\mathsf{G}_{[\,0,\,N\,]}\bigwedge_{j=1}^2(\nu_j^{\mathrm{O}}, 1) \!\rightarrow\! \mathsf{F}_{[\,0,\,N-3\,]}\mathsf{G}_{[\,0,\,3\,]} \bigwedge_{j=1}^2 (\nu_j^{\mathrm{E}}, 0.05))$. 
Using Eq.~\eqref{eq:moments}, we convert the \gls{dstl} formula $\varphi$ into an \gls{stl} formula.

\subsection{Experimental Configuration of Case II}

Case II considers a deterministic \gls{ev} modeled by Eq.~\eqref{eq:ego} and a stochastic \gls{ov} modeled by Eq.~\eqref{eq:linear_dy_compact_op}. The details of the models are given Appx.-B and Appx.-C. We use a deterministic variable $y_k\!\in\!\mathbb{R}^4$ and a stochastic variable $w_k\!\in\!\mathbb{R}^4$ to represent the outputs of the \gls{ev} and the \gls{ov}, respectively. Besides, we use another stochastic \gls{ov} model to denote the dynamics of a group of pedestrians, with a stochastic output variable $p_k\!\in\!\mathbb{R}^4$. The elements of these variables, i,e., $y_{i,k}$, $w_{i,k}$, and $p_{i,k}$ for $i\!\in\!\{1,2,3,4\}$ denote the longitudinal positions, the latitudinal positions, the orientations, and the linear velocities of the \gls{ev}, the \gls{ov}, and the pedestrian, respectively.

Following the natural language description in Sec.~\ref{sec:intersec}, the turning task is specified as \gls{dstl} predicates:
\textit{The \gls{ev} eventually reaches the target lane ($(\nu_1^{\mathrm{E}}, *):= |y_{1,k} \!-\! l/2| \!<\! 0.1\,$m$\,\wedge\, y_{2,k} \!>\! 3l/2$). Besides, the \gls{ev} should only turn after the first pedestrian crossing ($(\nu_2^{\mathrm{E}}, *): (y_{1,k}\!<\!-1.2l)\! \rightarrow\! (|y_{2,k}\!+\!l/2| \!<\! 0.1\,\mathrm{m})$) and should always avoid collisions with the \gls{ov} and the pedestrians at a risk level $0.05$ ($(\nu_1^{\mathrm{S}}, 0.05)\!:=\!P(|w_{1,k} \!-\! y_{1,k}| \!<\! d_{\mathrm{safe},1}^{\mathrm{O}}\wedge|w_{2,k} \!-\! y_{2,k}| \!<\! d_{\mathrm{safe}, 2}^{\mathrm{O}}) \!\geq\! 0.95$ and $(\nu_2^{\mathrm{S}}, 0.05)\!:=\!P(|p_{1,k} \!-\! y_{1,k}| \!<\! d_{\mathrm{safe},1}^{\mathrm{P}}\wedge|p_{2,k} \!-\! y_{2,k}| \!<\! d_{\mathrm{safe}, 2}^{\mathrm{P}}) \!\geq\! 0.95$). }
Here, $*$ denotes an arbitrary value in $[\,0,\,1\,]$, $l=8\,$m is the width of each lane, and $d_{\mathrm{safe},i}^{\mathrm{O}}\!=\!4\,$m and $d_{\mathrm{safe},i}^{\mathrm{P}}\!=\!2\,$m are safe distances between the \gls{ev} and the \gls{ov} and the pedestrians, respectively, for $i\!\in\!\{1,2\}$ representing the longitudinal and latitudinal directions, respectively.

The intention-aware \gls{dstl} specification for the task is $\varphi\!:=\!( \mathsf{G}_{[\,0,\,N\,]} \bigwedge_{j=1}^2 (\nu_j^{\mathrm{S}}, 0.05) )\wedge(\mathsf{F}_{[\,0,\,N-3\,]}\mathsf{G}_{[\,0,\,3\,]} \bigwedge_{j=1}^2 (\nu_j^{\mathrm{E}}, *))$. \\ The no-awareness one is $\varphi'\!:= \!(\mathsf{F}_{[\,0,\, N-3\,]}\mathsf{G}_{[\,0,\,3\,]}\! \bigwedge_{j=1}^2 (\nu_j^{\mathrm{E}}, *))$. Using Eq.~\eqref{eq:moments}, we convert these \gls{dstl} formulas into \gls{stl} formulas for the comparison study.

\bibliographystyle{IEEEtran}
\bibliography{IEEEabrv, reference.bib}

\balance

\end{document}